\input{psfig.sty}

\documentclass{aa}

\def\gta{\ga}
\def\1636{4U~1636--53}

\begin{document}

   \thesaurus{06     
              (03.11.1;  
               16.06.1;  
               19.06.1;  
               19.37.1;  
               19.53.1;  
               19.63.1)} 
\title{On the Correlated Spectral and Timing Properties of \1636: an 
Atoll Source at High Accretion Rates}

  \subtitle{}

   \author{Tiziana Di Salvo \inst{1} 
   \and Mariano M\'endez \inst{2} \and Michiel van der Klis \inst{1}}
   
   \offprints{T. Di Salvo}

   \institute{Astronomical Institute "Anton Pannekoek," University of 
              Amsterdam and Center for High-Energy Astrophysics,
              Kruislaan 403, NL 1098 SJ Amsterdam, the Netherlands\\
              email: disalvo@science.uva.nl, michiel@science.uva.nl
         \and
             SRON National Institute for Space Research, Sorbonnelaan 2, 
             3584 CA Utrecht, the Netherlands\\
             email:  M.Mendez@sron.nl}

   \date{Received ; accepted }
   
   \authorrunning{Di Salvo, M\'endez, \& van der Klis}
   \titlerunning{Spectral and Timing Properties of \1636}

   \maketitle

\begin{abstract}

We analyze $\sim 600$ ks data from the X-ray observatory Rossi X-ray Timing
Explorer of the neutron star low mass X-ray binary and atoll source \1636.
These observations span almost three years, from April 1996 to February
1999. The color-color and hardness-intensity diagrams 
show significant secular shifts of the atoll track, similar to
what is observed for some Z sources. These are most evident in the 
hardness-intensity diagram, where shifts in intensity up to $\sim 20\%$
are observed.  We find that the intensity shifts in the    
hardness-intensity diagram are responsible for the parallel tracks observed 
in the kilohertz quasi periodic oscillations frequency vs.\ intensity diagram.  
While the parallel tracks in the frequency vs.\ color plane partially overlap, 
systematic long term shifts are still evident.
We also study the broad band power spectra of \1636 as
a function of the source position $S_{\rm a}$ along the track
in the color-color diagram. These power spectra are all characteristic
of the lower and upper banana state in atoll sources, showing very low 
frequency noise, band limited noise, and sometimes one or two kHz QPOs.
We find that the very low frequency noise in some intervals is not well
described by a power law because of an excess of power between 20 and 30 mHz,
which can be fitted by a Lorentzian.  Also, the characteristic frequency
of the band limited noise shows a trend to decrease with increasing 
$S_{\rm a}$ at high $S_{\rm a}$ values.


\end{abstract}

\section{Introduction}

Low Mass X-ray Binaries (LMXB) containing weakly magnetized neutron stars 
can be divided into two classes based on their correlated X-ray spectral 
and timing behavior: the so-called Z and atoll sources  
(Hasinger \& van der Klis 1989).  
Z sources are thus named as they trace out roughly Z-shaped tracks in X-ray 
color-color or hardness-intensity diagrams on a time scale of hours to days. 
They are among the most luminous LMXBs, with X-ray luminosities 
close to the Eddington limit, $L_{\rm Edd}$. Atoll sources sometimes produce
patterns which are reminiscent of a `C', have luminosities in the range 
0.01--0.5 $L_{\rm Edd}$, and often show X-ray bursts. 
Their motion in the diagram is slower (on time scales of days to weeks)
in the upper left parts of the diagram, the so-called islands, 
while it is faster (on time scales of hours to a day) in the bottom and 
right-hand parts of the track, called lower and upper banana, respectively
(see e.g.\ Prins \& van der Klis 1997, M\'endez et al.\ 1997, Di Salvo et al.\ 
2001; see also Muno et al.\ 2002, Gierlinski \& Done 2002,
van Straaten et al.\ 2002a, Barret \& Olive 2002, for recent work on the 
behavior of atoll sources in the color-color diagram). In a given source, 
the islands usually correspond to lower flux levels than the banana branch.  

There is evidence that most timing and spectral characteristics of 
these sources depend in a simple way on the position along the Z or atoll 
track (see van der Klis 1995, 2000 for reviews). 
In fact, the characteristic strengths and frequencies of the timing 
variability of a source usually vary in a monotonic way as the source moves 
along the track, although some sources can show more complicated behaviors 
(see, e.g., the case of GX 17+2, Homan et al.\ 2002).
This general and regular behavior suggests that a single parameter, usually 
proposed to be the mass accretion rate, $\dot M$, governs most of this 
phenomenology; changes in $\dot M$ probably cause changes in the 
accretion flow, which simultaneously affect the X-ray spectrum and the rapid 
X-ray variability.
In this scenario the mass accretion rate is thought to
increase from the horizontal branch to the flaring branch in Z sources, and
from the islands to the banana branch in atoll sources (see e.g. Hasinger
et al.\ 1990, van der Klis 1995). 

At low inferred mass accretion rate the power spectra of these sources are 
usually dominated by band limited noise (BLN) components. 
This noise component, usually called in the literature high-frequency noise
in atoll sources and low-frequency noise in Z sources, has a flat shape up 
to a break frequency, $\nu_{\rm break}$, above which it decreases with 
increasing frequency; it has been often described by a broken power law or 
a power law with a cutoff at high frequencies.  However, since recently, 
a Lorentzian model for this component is preferred (Olive et al.\ 1998; 
Belloni et al.\ 2002; van Straaten et al.\ 2002b), 
since this facilitates the comparison with the power spectra 
of black hole candidate binaries, which are well fitted by Lorentzians 
(see e.g.\ Miyamoto et al.\ 1991; Nowak 2000), and because a Lorentzian gives 
sometimes a better description of the noise shape than a broken power law 
(van Straaten et al.\ 2002b). 
Furthermore, Lorentzians are more suitable than power laws to describe
components that show large changes in quality factor, which have been
observed in some sources (see e.g.\ Di Salvo et al.\ 2001; van Straaten 
et al.\ 2002b).  
At high inferred mass accretion rate another type of noise component, 
with a power-law shape, $P(\nu) \propto \nu^{-\alpha}$ with index $\alpha 
\sim 1.5-2.0$, called Very Low Frequency Noise (VLFN), dominates at  
frequencies less than $\sim 1$ Hz.
The properties of both these noise components strongly correlate with the 
position of the source in the color-color diagram.  At the lowest inferred 
$\dot M$, the BLN is quite strong (with a fractional rms amplitude of 
10--20\%). When the mass accretion rate increases, as inferred from the source 
position in the color-color diagram, the rms amplitude of the BLN decreases 
to $\le 2\%$. 
Simultaneously, the break frequency increases from a few Hz to tens of Hz
(this occurs at least in the part of the color-color diagram in which the
kHz QPOs are detected; see e.g.\ Wijnands \& van der Klis 1999; Psaltis 
et al.\ 1999a, Di Salvo et al.\ 2001, and references therein). 
At the highest inferred $\dot M$ levels, when the kHz QPOs are no longer
detected, a BLN component is sometimes still detected, usually with a
characteristic frequency of $10-20$ Hz (e.g.\ Hasinger \& van der Klis 1989,
Di Salvo et al.\ 2001, van Straaten et al.\ 2002b, see also Sect. 5).
The VLFN, on the other hand, usually has its lowest fractional amplitude at 
low inferred accretion rate ($\le 1\%$) and its amplitude gradually increases 
with inferred accretion rate, sometimes up to $\sim 5\%$ (see van der Klis 
1995 for a review).

The BLN in atoll and Z sources is usually accompanied by Quasi-Periodic 
Oscillations (QPOs) at  
frequencies between 10 and 40~Hz, slightly higher than the break frequency of 
the BLN. These are the so-called Low-Frequency QPO (LFQPO) in atoll sources,
and horizontal branch oscillations (HBO) in Z sources.  Similar to the break 
frequency of the BLN, the frequency of these QPOs increases when the 
source moves in the color-color diagram in the sense of increasing mass 
accretion rate (e.g.\ van der Klis et al.\ 1985; Stella \& Vietri 1998;
Ford \& van der Klis 1998; van Straaten et al.\ 2000; Di Salvo et al.\ 2001;
Psaltis et al.\ 1999a, and references therein).
The rms amplitude of the QPOs decreases more or less gradually when the 
source moves from the horizontal branch to the normal branch in Z sources, 
and from the islands to the banana in atoll sources.

Observations with the large area instrument on board the Rossi X-ray Timing 
Explorer (RXTE) led to the discovery in 1996 of QPOs at kilohertz frequencies 
(kHz QPOs) in the persistent emission of Z and atoll sources. 
To date, kHz QPOs have been observed in the power spectra of more than 20 
LMXBs, with frequencies ranging from a few hundred Hz up to $1200-1300$ Hz 
(the maximum kHz QPO frequency of 1330 Hz has been measured in 4U 0614+09,
van Straaten et al.\ 2000; see van der Klis 2000 for a review).  
Usually two kHz QPO peaks (``twin peaks'') are simultaneously observed, the 
difference between their centroid frequencies being in the range 250--350~Hz.
The kHz QPO frequencies increase when the source moves in the color-color 
diagram in the sense of increasing inferred $\dot M$.  However, while
there is a general correlation between kHz QPO frequencies and position
in the color-color diagram, the correlation between kHz QPO frequencies
and the source X-ray count rate or X-ray flux in the 2--50~keV energy range
is complex; a correlation is observed on short time scales (hours to days), 
but not on longer time scales. Therefore, when observed at different 
epochs, in a frequency vs.\ flux diagram a source produces different tracks 
that are approximately parallel (e.g.\ M\'endez et al.\ 1999). 
A possible explanation of this behavior is that most 
timing and spectral parameters are determined by the dynamical properties of 
the accretion disk, while the X-ray flux is determined by the total accretion
rate, which may be different from the instantaneous accretion rate through 
the disk if, for instance, matter can flow radially close to the neutron star 
(see e.g.\ van der Klis 2001; note, however, that the radial flow cannot be 
completely independent of the disk flow, see M\'endez et al.\ 2001).  
This might also explain the lack of a simple correlation between the X-ray 
flux (which is usually considered to be a good measure of $\dot M$) and other
timing properties and/or position in the color-color diagram on time scales 
of days to weeks (see e.g. M\'endez \& van der Klis 1999, Ford et al.\ 2000, 
and references therein).
The whole scenario is not clear yet, and in this paper we will continue 
to use the widely adopted terminology of ``inferred mass accretion rate'' to 
indicate the position of the source in the color-color diagram, keeping in 
mind that, while this terminology might be correct for the short term 
(hours to a day), in the longer term there might not be an exact 
correspondence between instantaneous accretion rate and position in the 
color-color diagram.

Quasi-coherent oscillations during type-I X-ray bursts have now been 
detected in several sources, and are all in the rather narrow 
frequency range between 300 and 600~Hz (see van der Klis 2000;
Strohmayer 2001 for reviews).
Usually, the frequency of burst oscillations increases by 1--2\,Hz during 
the tail of the burst, and saturates to an ``asymptotic frequency'' which 
in a given source is consistent with being constant. This asymptotic 
frequency is interpreted as the neutron star spin frequency, due to a hot spot 
(or spots) in an atmospheric layer of the rotating neutron star.
For those sources in which both burst oscillations and kHz QPOs have 
been observed, it has been found that the frequency separation
between the two simultaneous kHz QPOs is similar to the frequency of the
burst oscillations, or half that value. 
This suggests a beat frequency mechanism as the origin
of the kHz QPOs (Strohmayer et al.\ 1996; Miller et al.\ 1998), 
where the higher-frequency kHz QPO (upper peak) is interpreted as the 
Keplerian frequency at the innermost edge of the accretion disk, and 
the lower-frequency kHz QPO (lower peak) is the beat between this Keplerian 
frequency and the spin frequency, $\nu_{\rm s}$, of the neutron star. 
However, the peak separation is sometimes significantly smaller
than the frequency of the burst oscillations (e.g.\ M\'endez et al.\ 
1998a), and decreases as the frequency of the kHz QPOs
increases (van der Klis et al.\ 1997; M\'endez et al.\ 1998a; 
M\'endez \& van der Klis 1999).  This led to modifications to
the simple beat frequency model (Lamb \& Miller 2001), or to 
different models for the kHz QPOs (see e.g.\ Stella \& Vietri 1999;
Osherovich \& Titarchuk 1999; Klu\'zniak \& Abramowicz 2002).

4U 1636--536 is one of the most interesting LMXBs of the atoll class. It 
contains a neutron star   
accreting matter from a companion star of mass $\sim 0.4$ $M_\odot$  
with an orbital period of 3.8 hr (see, e.g., van Paradijs et al.\ 1990).
Type-I X-ray bursts were observed several times from this source, sometimes
with unusual shapes (e.g.\ Turner \& Breedon 1984; van Paradijs et al.\ 1986),
or unusually long durations (e.g.\ Wijnands 2001).
Duration and temperature of the X-ray bursts in this source strongly
correlate with the X-ray spectral and fast variability characteristics 
of the persistent emission, implying a correlation of all these 
characteristics with the mass accretion rate or disk dynamical properties
(van der Klis et al.\ 1990).  
Recently highly coherent oscillations at a frequency of $\sim 582$ Hz
have been observed for $\sim 800$ s during a ``superburst'' from this
source (Strohmayer \& Markwardt 2002). No evidences of higher harmonics
or the subharmonic at 290 Hz have been found. The high coherence of
this signal provides further support to a connection between burst
oscillations and spin of the neutron star. However, the frequency measured
during the superburst is significantly higher than any of the asymptotic
burst oscillation frequencies measured for \1636 ($\sim 581$ Hz, see e.g.\
Strohmayer et al.\ 1998; Giles et al.\ 2002).
Two simultaneous kHz QPOs have been observed in this source, with rms 
amplitudes increasing with photon energy (Zhang et al.\ 1996; Wijnands et al.\ 
1997a).  The frequency of the lower kHz QPO increases, and its 
amplitude decreases, with inferred $\dot M$, and at high $\dot M$ the frequency 
difference between the two QPOs is in the range $240-280$ Hz, significantly 
lower than half the frequency of the oscillations detected during type-I 
bursts (M\'endez et al.\ 1998a). 
Recently it has been shown that, at low inferred mass accretion rate,
the kHz QPO peak separation in \1636 is $\sim 323$ Hz, which exceeds  
the neutron star spin frequency (or half its value) as inferred from burst 
oscillations (Jonker et al.\ 2002a). 

Third, and possibly fourth, weaker kHz QPOs have been discovered
simultaneously with the previously known kHz QPO pair (Jonker et al.\ 
2000); the new kHz QPOs are at frequencies of $\sim 58$ Hz above and below, 
respectively, the frequency of the lower kHz QPO, suggesting that these are 
sidebands to the lower kHz QPO. 
\1636, together with 4U~1608--52 and Aql~X--1, also shows QPOs at mHz 
frequencies (Revnivtsev et al.\ 2001).  This feature seems to occur only in 
a rather narrow range of mass accretion rates, corresponding to X-ray 
luminosities of $\sim (0.5-1.5) \times 10^{37}$ ergs/s, and they disappear 
after X-ray bursts. Contrary to the general behavior of QPOs, the rms 
amplitude of the mHz QPOs strongly decreases with energy.

A systematic study of the correlated rapid X-ray variability and X-ray 
spectral properties of \1636, as derived from color-color and 
hardness-intensity diagrams, was performed on EXOSAT data by Prins \& 
van der Klis (1997).  They found that over a period of two years the source 
traced out similar patterns in the color-color and hardness-intensity diagrams. 
However, in observations taken months apart the position of the patterns was
slightly (a few percent) shifted in the color-color plane, corresponding
to much larger differences in the intensity. 
They also found that the power spectral components were clearly correlated 
to the position of the source in this pattern corrected for the shifts due
to the secular motion; in particular, with increasing inferred mass accretion 
rate the fractional rms amplitude of the power-law shaped VLFN increased, 
while the BLN component amplitude, as well as its cutoff frequency, decreased. 

In this paper we present the broad band (0.01-2048 Hz) power 
spectra of \1636 extracted at different positions of the source in the 
color-color diagram, as well as a study of the secular variations of 
the position of the atoll pattern in both the hardness-intensity and the 
color-color diagram.

\section{Observations}

We analyzed all the observations from the public RXTE archive performed 
between April 1996 and February 1999.  The log of these observations is 
presented in Table~\ref{tab1}.  
We used data from the Proportional Counter Array (PCA; Zhang et al.\ 1993) 
on board RXTE, which consists of five co-aligned Proportional Counter 
Units (PCU), sensitive in the energy range $2-60$ keV, with a total 
collecting area of 6250 cm$^2$ and a field of view, delimited by collimators, 
of $1^\circ$ FWHM.  We selected intervals for which the
elevation angle of the source above the Earth limb was greater
than 10 degrees. 
Several bursts occurred during these observations; we excluded those
intervals ($\sim 500$ s around each burst) from our analysis.  
We used the Standard 2 mode data (16 s time resolution) to produce 
light curves and the color-color diagram, and Event data, with time 
resolution of 1/8192 s ($\sim 125 \mu$s), to produce Fourier 
power spectra.  For the power spectra we also discarded those data 
($\sim 8$\% of the total) where one or more of the five PCUs were 
switched off.
\begin{table}
\caption[]{RXTE observations of \1636 used in this paper.
The count rate has been normalized to 3 PCA units and is corrected for the 
background.
}
\scriptsize
\label{tab1}
\begin{tabular}{c|c|c|c} 
\hline \hline
Observation & Start Time  & Total Time & Averaged Rate   \\ 
 	    & (UTC)       & (sec)      &  (c/s)          \\ 
\hline
10072-03-03-00 & 23-04-1996 03:05 & 1632 & 1351.16 \\
10072-03-02-00 & 29-05-1996 01:13 & 5552 & 1030.66 \\
10088-01-01-00 & 27-04-1996 13:45 & 8928 & 1040.03 \\
10088-01-04-00 & 27-04-1996 17:02 & 3296 & 1099.83 \\
10088-01-05-00 & 28-04-1996 12:09 & 8576 & 1216.73 \\
10088-01-02-00 & 29-04-1996 17:37 & 6256 & 1043.76 \\
10088-01-03-00 & 30-04-1996 16:03 & 6848 & 1052.10 \\
10088-01-07-01 & 09-11-1996 20:53 &12210 & 1279.49 \\ 
10088-01-07-00 & 14-11-1996 18:12 &19790 & 1802.72 \\
10088-01-07-02 & 28-12-1996 22:25 &13790 & 1251.22 \\
10088-01-08-00 & 29-12-1996 03:45 &19170 & 1452.39 \\
10088-01-08-010& 29-12-1996 16:32 &22670 & 1286.36 \\
10088-01-08-01 & 29-12-1996 23:27 &10020 & 1224.70 \\
10088-01-08-02 & 30-12-1996 03:24 &25260 & 1405.65 \\
10088-01-08-04 & 31-12-1996 01:09 &16380 & 1396.83 \\
10088-01-08-030& 31-12-1996 10:47 &25550 & 1279.06 \\  
10088-01-08-03 & 31-12-1996 18:26 & 9792 & 1291.42 \\
10088-01-06-010& 05-01-1997 22:05 &27390 & 1310.02 \\
10088-01-06-01 & 06-01-1997 06:01 & 3008 & 1292.10 \\
10088-01-06-07 & 06-01-1997 08:33 & 9632 & 1306.41 \\
10088-01-06-02 & 07-01-1997 03:28 &13500 & 1674.33 \\
10088-01-06-03 & 07-01-1997 08:05 &12020 & 1642.56 \\
10088-01-06-04 & 08-01-1997 04:32 &15340 & 1412.60 \\
10088-01-06-06 & 08-01-1997 09:37 & 6448 & 1434.10 \\
10088-01-06-000& 08-01-1997 22:14 &21010 & 2026.24 \\
10088-01-06-00 & 09-01-1997 04:22 &14850 & 1821.97 \\
10088-01-06-05 & 09-01-1997 20:37 &12210 & 1701.56 \\ 
10088-01-06-08 & 10-01-1997 00:02 &18990 & 1826.38 \\ 
10088-01-09-00 & 22-02-1997 06:37 &19760 & 1651.99 \\
10088-01-09-01 & 23-02-1997 04:49 &26270 & 1865.01 \\
10088-01-09-02 & 24-02-1997 04:46 &20640 & 1581.61 \\
30053-02-01-000& 24-02-1998 23:26 &22690 &  976.50 \\
30053-02-01-001& 25-02-1998 05:58 &25090 & 1037.83 \\
30053-02-01-00 & 25-02-1998 13:15 & 8464 & 1187.74 \\
30053-02-02-02 & 19-08-1998 08:15 &15150 & 1026.38 \\ 
30053-02-02-01 & 19-08-1998 13:03 &15150 & 1059.31 \\ 
30053-02-01-03 & 19-08-1998 17:50 & 3648 & 1059.37 \\
30053-02-01-01 & 20-08-1998 01:50 & 1648 & 1027.12 \\
30053-02-01-02 & 20-08-1998 03:26 & 2064 & 1075.32 \\
30053-02-02-00 & 20-08-1998 05:11 &26110 & 1241.44 \\
30056-03-01-00 & 29-12-1998 15:35 & 3936 & 1548.72 \\
30056-03-01-03 & 30-12-1998 14:41 & 1424 & 1861.03 \\
30056-03-01-01 & 30-12-1998 15:34 & 4032 & 1854.93 \\
30056-03-01-04 & 31-12-1998 14:41 & 1472 & 1316.05 \\
30056-03-01-02 & 31-12-1998 15:33 & 4080 & 1479.34 \\
40028-01-01-00 & 12-01-1999 09:28 &21500 & 1521.64 \\
40028-01-02-00 & 27-02-1999 05:57 & 4576 & 1134.68 \\
\hline
\end{tabular}
\end{table}

\section{Color-Color and Hardness-Intensity Diagrams}

\begin{figure*}
\hbox{\hspace{2.5cm}\psfig{figure=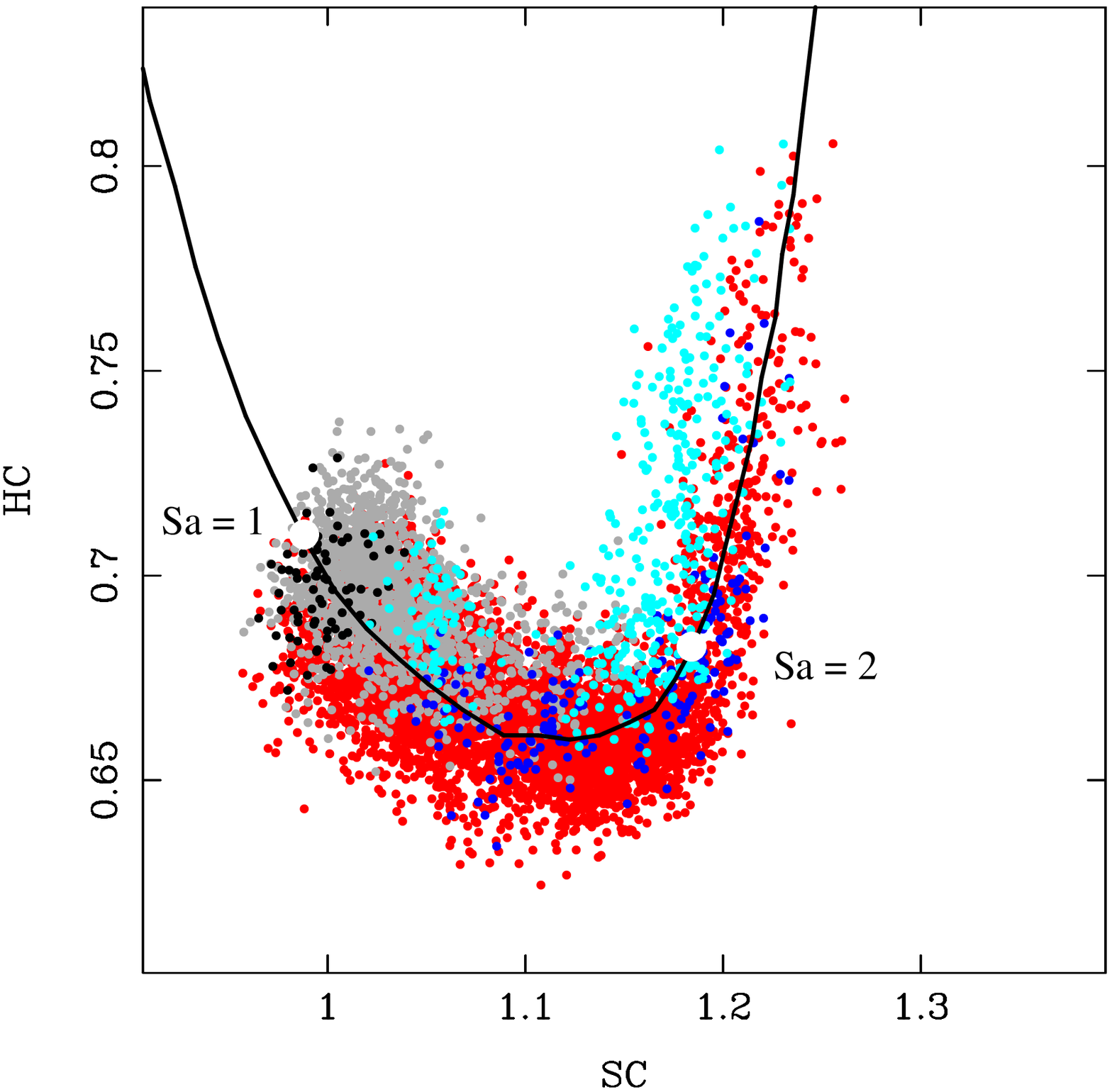,width=6.0cm}\hspace{1cm}
\psfig{figure=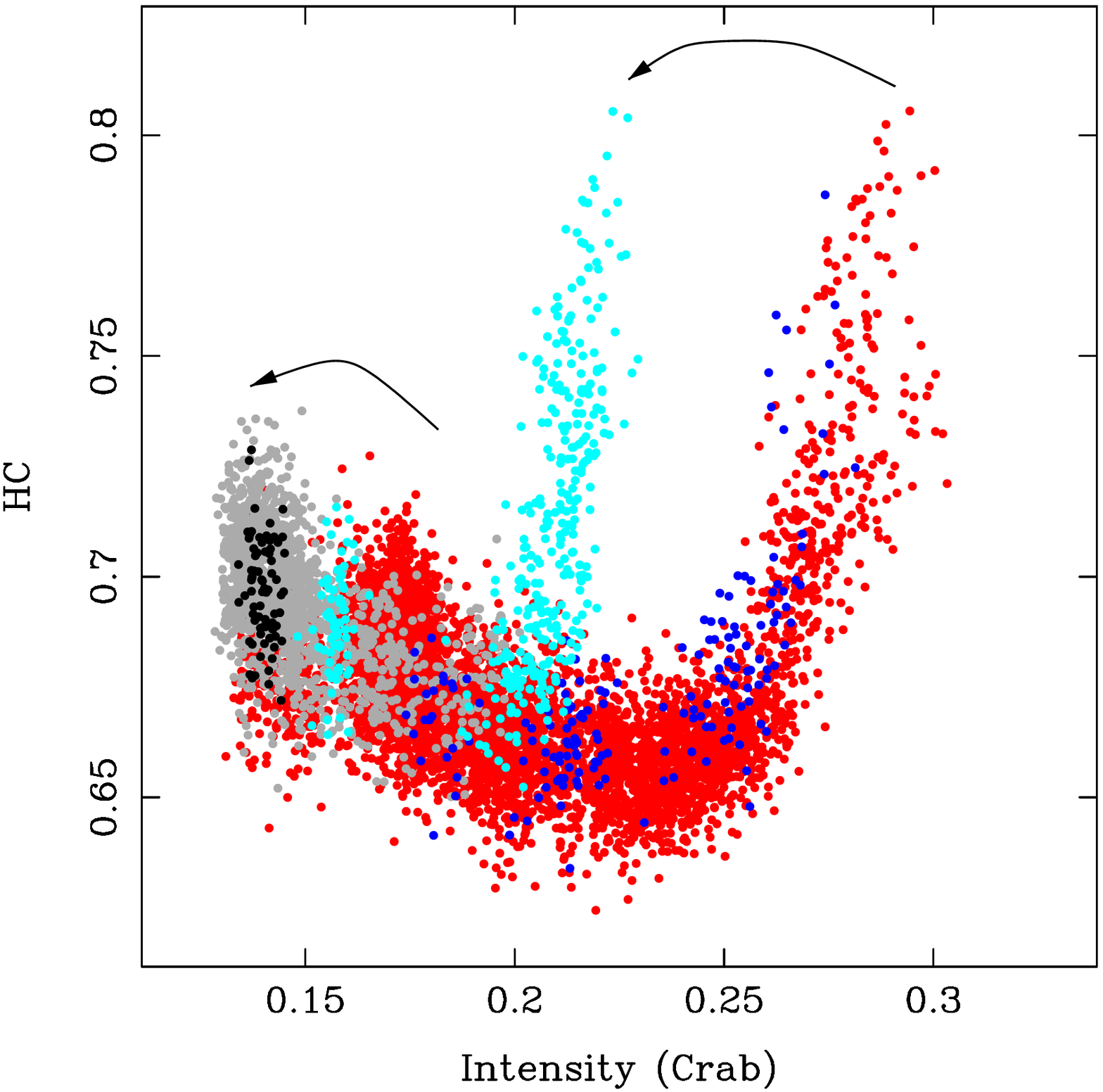,width=6.0cm}}
\hfill      \parbox[b]{18cm}{\caption[]{Color-color diagram (left) and 
hardness-intensity diagram (right) of \1636.  Each point represents 64 s of 
data. The soft and hard colors are defined as the ratio of the count rate
in the bands 3.5--6.4 keV/2.0--3.5 keV and 9.7--16 keV/6.4--9.7 keV, 
respectively. The intensity is defined as the source count rate in the
energy band 2--16 keV.  The soft and hard colors, as well as the source 
intensity, are normalized to the Crab values.
Note that 1 Crab is $\sim 2.56 \times 10^{-9}$ ergs cm$^{-2}$ s$^{-1}$ 
keV$^{-1}$, corresponding to an average $2-16$ keV count rate of 
$\sim 2430$ c/s/PCU. 
Different colors indicate datasets from different observations 
(P10072: black; P10088: red; P30053: gray; P30056: blue; P40028: light blue; 
see Table~\ref{tab1}). The black curve in the color-color diagram is the 
spline that we defined to parametrize the position of the source as a 
function of the coordinate $S_{\rm a}$, and the two big white points indicate 
the places for which we defined the values $S_{\rm a} = 1$ at ($HC = 0.99$, 
$SC = 0.71$) and $S_{\rm a} = 2$ at ($HC = 1.18$, $SC = 0.68$), along the 
spline (see text).
The arrows in the hardness-intensity diagram indicate those parts of the 
diagram where the shifts are particularly evident.
}\label{fig:fig1}}
\hbox{\hspace{3cm}\psfig{figure=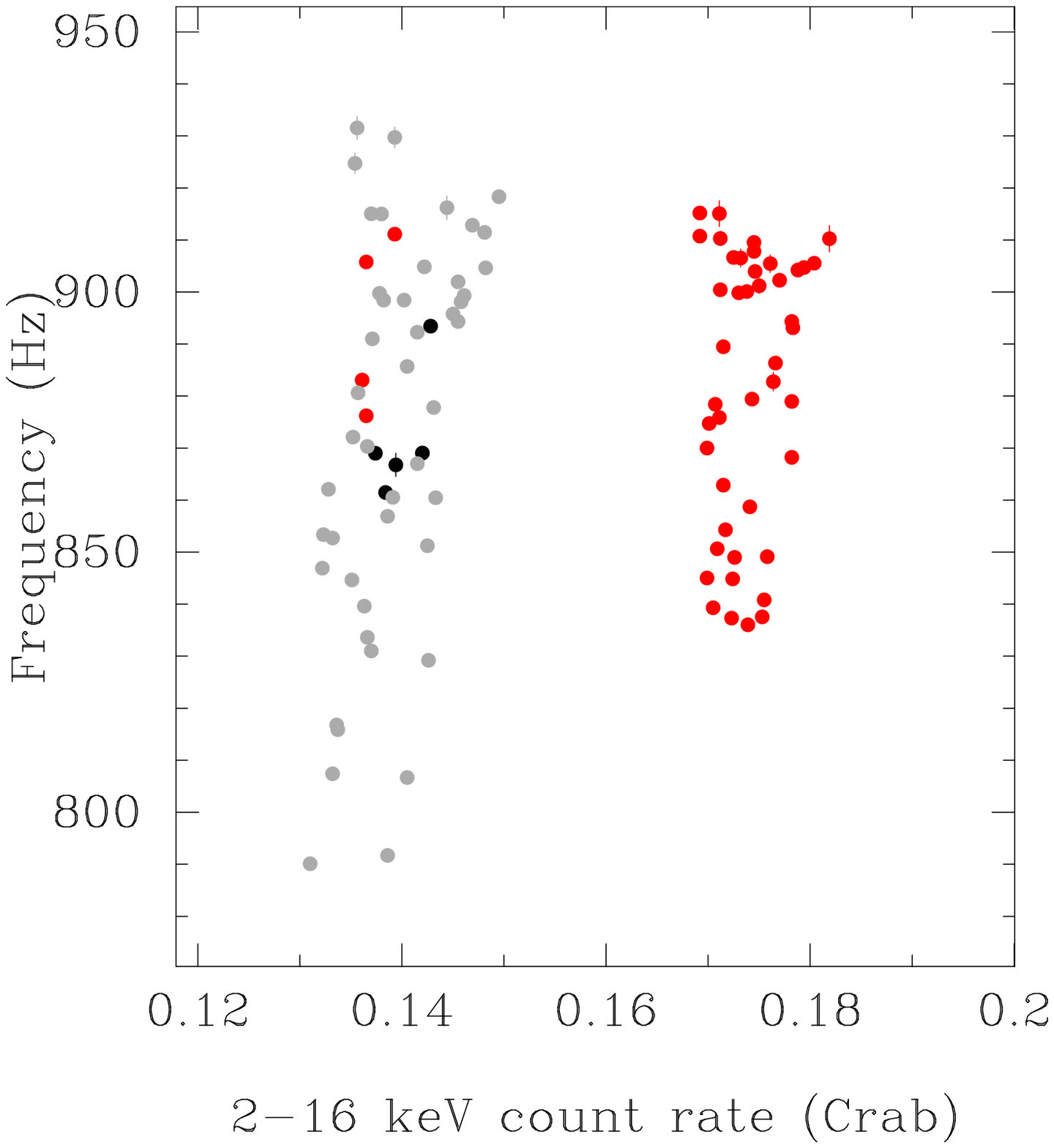,width=5.5cm}\hspace{1cm}
\psfig{figure=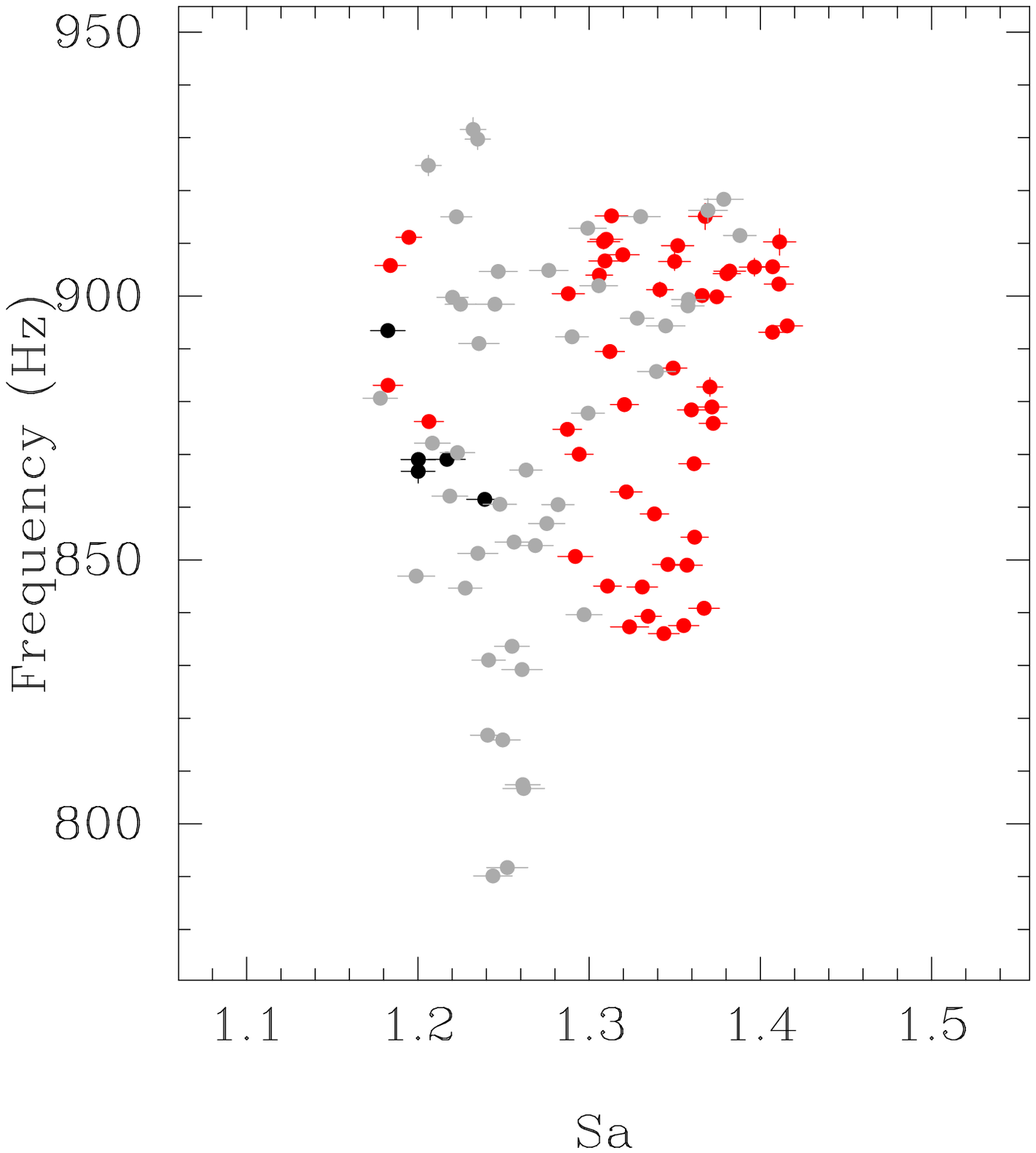,width=5.5cm}}
\vskip 0.5cm
\hbox{\hspace{3cm}\psfig{figure=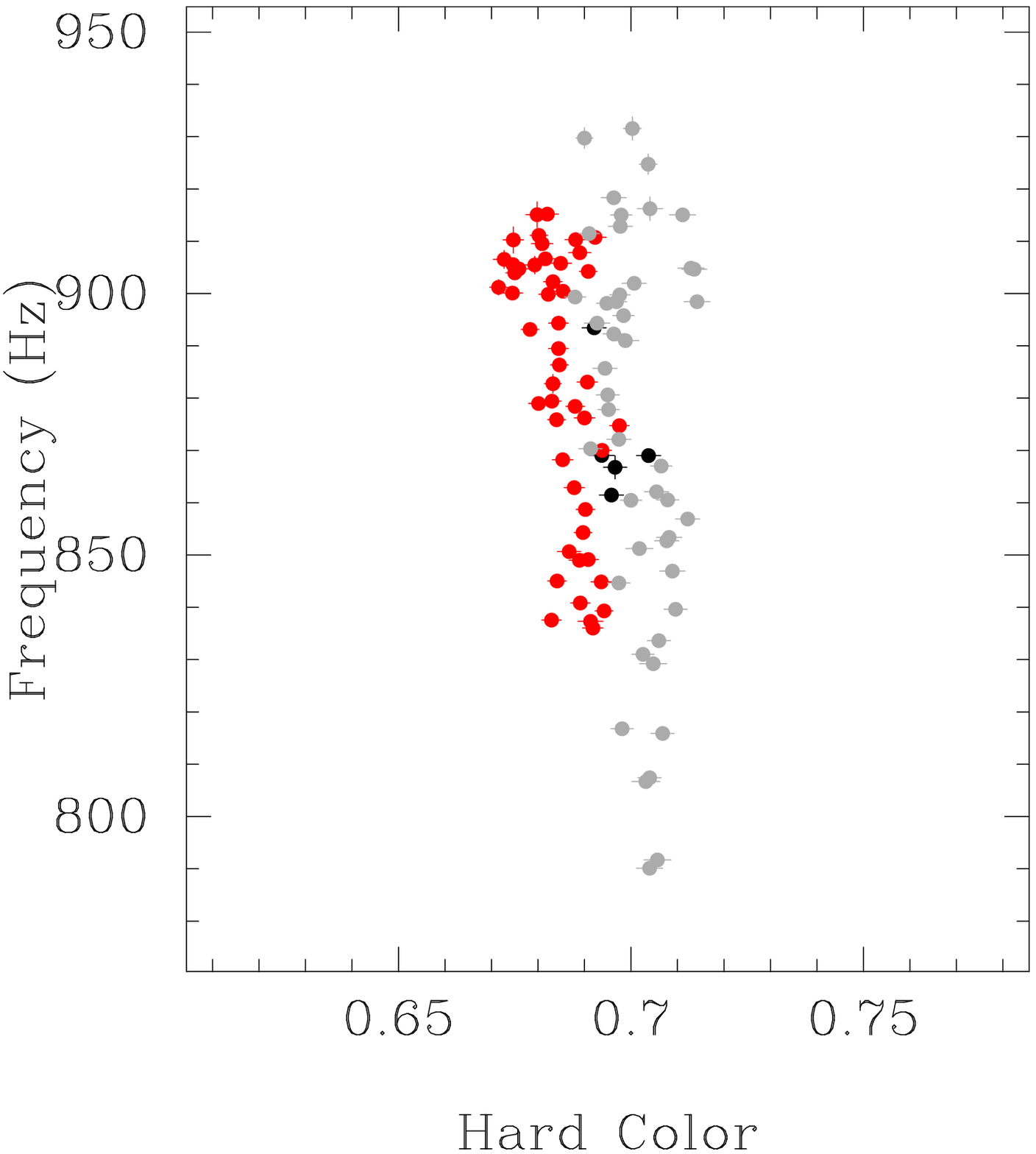,width=5.5cm}\hspace{1cm}
\psfig{figure=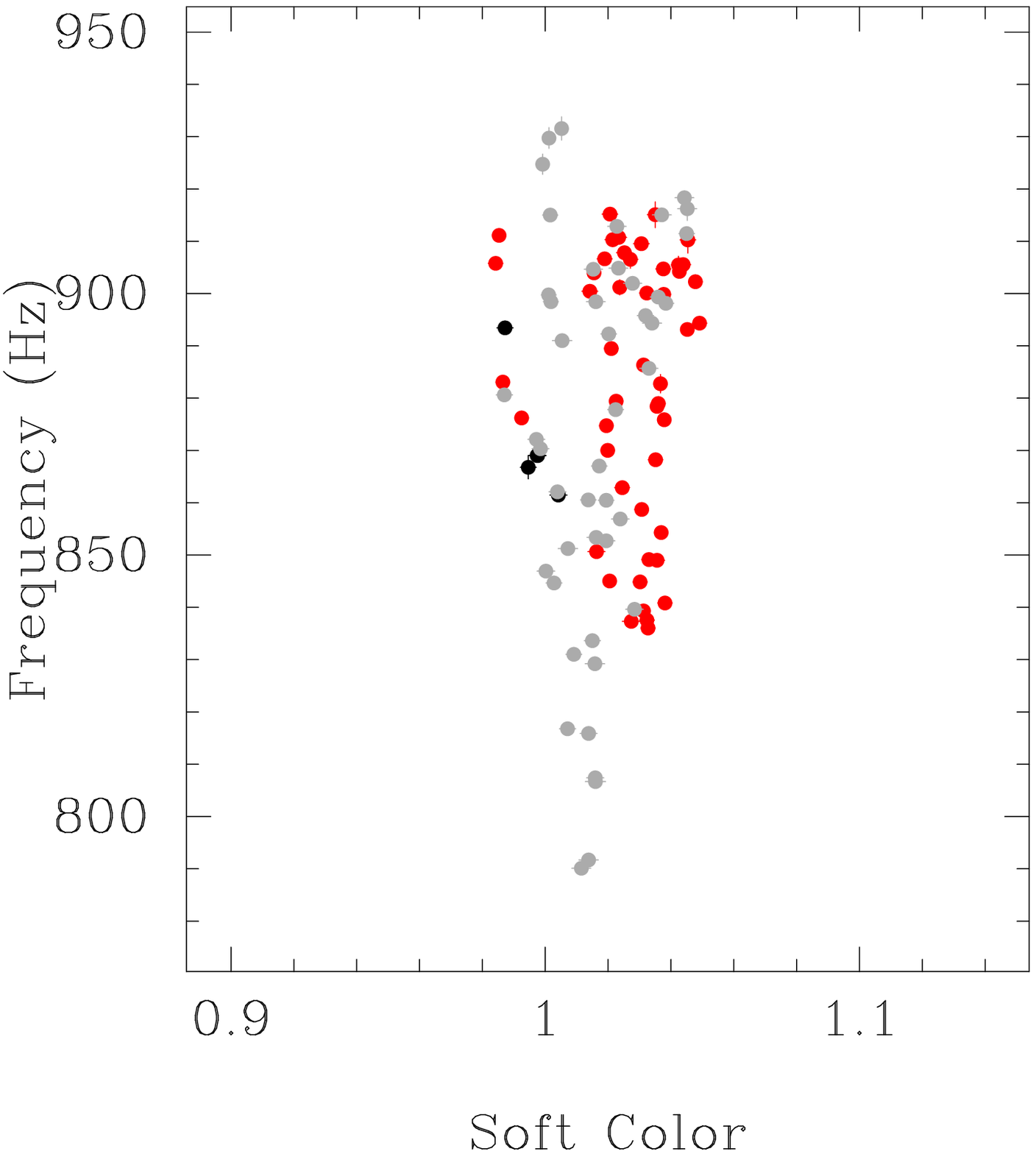,width=5.5cm}}
\hfill      \parbox[b]{18cm}{\caption[]{The lower kHz QPO frequency in \1636 
plotted vs.\ the 2--16 keV source count rate in Crab units (upper left panel),
vs.\ $S_{\rm a}$ (upper right panel), vs.\ the hard color (lower left panel) 
and vs.\ the soft color (lower right panel).  
Different colors indicate different datasets;
P10088 (from 1996 April 27 to 1996 April 30, red points on the low-flux track), 
P10072 (1996 May 29, black points), 
P10088 (from 1996 November 9 to 1997 January 6, red points on the high-flux 
track), P30053 (1998 February 24--25 and 1998 August 19--20, gray points). 
}\label{fig:fig2}}
\end{figure*}

We produced background subtracted lightcurves binned at 16~s using
PCA Standard 2 mode data from PCUs 0, 1, and 2, and the PCA background 
model 2.1b.  The lightcurves were divided into four energy bands, 
$2.0-3.5$, $3.5-6.4$, $6.4-9.7$, $9.7-16.0$ keV (according to the 
correspondence between fixed energy channels and energy valid for 
the PCA gain epoch 3).  We defined the soft color as the
ratio of the count rate in the bands $3.5-6.4$ keV and $2.0-3.5$ keV, the
hard color as the ratio of count rate in the bands $9.7-16.0$ keV and
$6.4-9.7$ keV, and the intensity as the count rate in the energy band 
2--16 keV, in order to produce color-color and hardness-intensity diagrams.  
The observations considered here span almost three years, from 1996 April 23 
to 1999 February 27. All of them fall in the PCA gain epoch 3 (from 1996 
April 15 to 1999 March 22).
However, to take into account the (smaller) continuous gain changes 
occurring within one epoch, we used the same procedure already applied
in previous works (see e.g.\ Kuulkers et al.\ 1994, van Straaten et al.\ 2000, 
Di Salvo et al.\ 2001): we selected RXTE
observations of Crab obtained close to the dates of our observations,
and calculated the corresponding X-ray colors of Crab.  
Because the spectrum of Crab is supposed to be constant, any changes 
observed in the colors of Crab are most probably caused by changes in the 
instrumental response.  
Between April 1996 and February 1997 there are no significant shifts
in the Crab colors. Between February 1997 and February 1998
the soft and hard colors of Crab increased by $\sim 4$\,\%
and $\sim 2$\,\%, respectively. From February 1998 to February 1999
the Crab soft color slightly increased by up to $\sim 2.6\%$, while the hard
color remained almost constant (changes were less than $\sim 1\%$).
To take into account these variations we divided the colors calculated
for \1636 by the corresponding colors of Crab; these latter were
on average: $SC = 2.01$ and $HC = 0.595$. 
The changes in the PCA gain, although small, might also affect the source
count rate. On average, the Crab count rate decreased from April 1996 to
February 1999 by up to $\sim 4.5\%$. So we applied a similar correction to 
the source intensity, which was normalized to the Crab count rate 
(the average 2--16 keV Crab count rate was $\sim 2430$ c/s/PCU).
The color-color and hardness-intensity diagrams, normalized to the 
Crab and rebinned so that each point corresponds to 64 s of data, are shown 
in Fig.~\ref{fig:fig1}.
We have checked the validity of the correction based on Crab 
by using spectral fits of \1636 observations and the corresponding response 
matrices (see e.g.\ Barret \& Olive 2002). We find that the correction 
factors obtained in this way are very similar to those calculated using Crab. 
They are $\sim 2\%$ larger in soft color and $\sim 0.2\%$ larger in hard 
color than the Crab-based correction factors, which are comparable with the 
1\% uncertainty in the PCA response matrix and with the statistical 
errors on the colors. 

During these observations the source was in the so-called banana state, 
at high inferred mass accretion rate.
As it is apparent from Fig.~\ref{fig:fig1}, significant shifts of the 
position of the atoll track are observed in the hardness-intensity diagram. 
The most evident shift (indicated by a big arrow in Fig.~\ref{fig:fig1}) 
is between the dataset of observation P10088 (between April 1996 and February 
1997, red points) and the dataset of observation P40028 (between January 1999 
and February 1999, light-blue points) on the right-hand side of the 
hardness-intensity diagram, corresponding to the upper banana.  
Probably a shift is also present within the dataset of observation P10088 
(in particular between data obtained before 1996 April 30, which correspond 
to a lower flux level, and data obtained after 1996 November, which correspond 
to a higher flux level) and between this latter dataset and the dataset of 
observations P10072 (in 1996 April and May, black points) and P30053 (in 1998 
February and August, gray points) on the left-hand side of the diagram 
(lower banana, indicated by a small arrow in Fig.~\ref{fig:fig1}).  
In both parts of the hardness-intensity diagram the shift in intensity is 
$\sim 20\%$.  
Indeed, it seems that two parallel atoll tracks are present, one shifted
$\sim 20\%$ towards lower intensity with respect to the other one.
In the color-color diagram, these two atoll tracks overlap each other.  
However, similar shifts as in the hardness-intensity diagram, although much 
less evident, are present.  
This is especially clear in the upper banana (on the right 
side of the atoll track), where it is possible to see that the points of 
observation P10088 (red points) have soft color systematically higher than 
the points of observation P40028 (light-blue points).

To parametrize the position of the source along the atoll track we used 
the $S$ parametrization method that is usually used for Z sources 
(introduced by Hasinger et al.\ 1990; Hertz et al.\ 1992, and then refined in 
successive works, see e.g.\ Homan et al.\ 2002 and references therein).
We projected the points in the color-color diagram onto a spline, $S_{\rm a}$, 
in which two vertices are placed as reference in the atoll track. We set 
$S_{\rm a} = 1$ at the color-color locus (0.99, 0.71) and $S_{\rm a} = 2$ at 
(1.18, 0.68), as indicated in Fig.~\ref{fig:fig1}.  
A difficulty of this method in the case of atoll sources when compared to 
Z sources is the lack of sharp corners in the atoll track, which makes the 
choice of the vertex points rather arbitrary.

\section{Study of the Broad Band Power Spectra}

\subsection{kHz QPOs}

One of the aims of this work is to study the timing properties of \1636
as a function of the source spectral state, as derived from the position in
the color-color diagram.  We begin with the study of the kHz QPO frequencies, 
dividing the PCA lightcurve into 64-s long segments for which we calculated 
Fourier power spectra up to a Nyquist frequency of 2048 Hz.  
For each of these segments we measured the average frequency of the kHz QPOs 
and the position in the color-color diagram using the corresponding value of 
$S_{\rm a}$.  
Following M\'endez et al.\ (2001), we produced a dynamic 
power spectrum for each observation, and we identified those power spectra 
that showed strong QPOs at frequencies $\ga 250$ Hz. 
When two QPOs were present in a power spectrum, we selected the one 
at lower frequency, $\nu_{\rm low}$, which was generally the strongest.
We fitted the power spectra with kHz QPOs in the range $\nu_{\rm low} -
100$ Hz to $\nu_{\rm low} + 100$ Hz, using a function consisting of a
constant plus one Lorentzian. In a few of these 64-s segments the QPO 
was not statistically significant (less than $3\sigma$, single trial). 
In these cases we averaged several contiguous 64 s segments, 
but always less than 20, to increase the statistical significance of the
detection. We did not combine more than 20 consecutive power spectra
to avoid systematic effects (e.g.\ an artificial broadening) in the 
measurements due to the variation of the QPO frequencies, which can change 
by a few tens of Hz within a few hundred seconds. We discarded those data 
for which this procedure did not reveal any significant kHz QPO. Note, 
however, that in this way we might also have discarded data with weak 
QPOs, that are not significantly detected in segments of 1280 s.

\begin{figure}
\hbox{\hspace{0cm}\psfig{figure=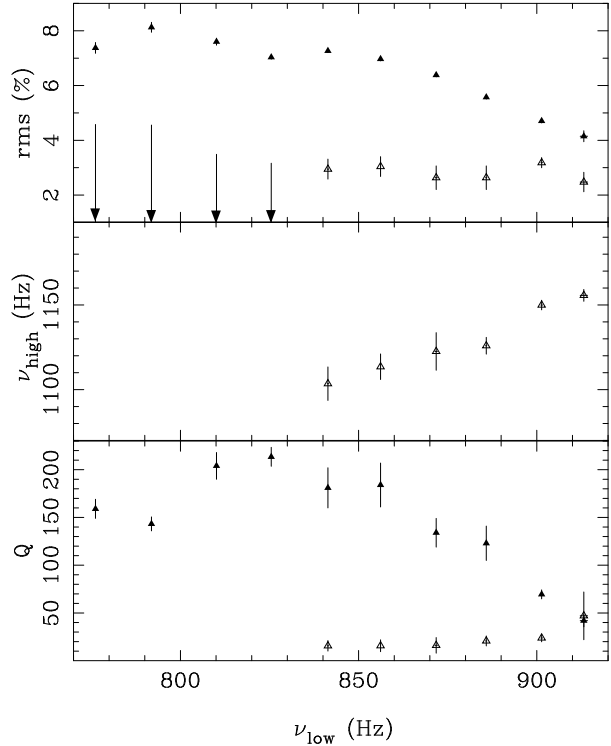,width=8.0cm}}
\caption[]{
Rms amplitude of the lower kHz QPO (filled triangles, top panel) and
the upper kHz QPO (open triangles, top panel), 
frequency of the upper kHz QPO (middle panel), 
quality factor, $Q$, of the lower kHz QPO (filled triangles, bottom panel)
and of the upper kHz QPO (open triangles, bottom panel)
of \1636 plotted against the frequency of the lower kHz QPO. 
For the first four intervals, where the upper kHz QPO was not detected with
high statistical significance, we fixed the upper kHz QPO frequency 
assuming a frequency separation 
between the two kHz QPOs of 260 Hz, and its quality factor assuming a typical 
value of 16. We then calculated the 90\% confidence level upper limits on its 
rms amplitude (which are shown in the top panel).}
\label{fig:fig3}
\end{figure}

In Fig.~\ref{fig:fig2} (left top panel) we plot the frequency of the kHz 
QPO we measured as a function of the source intensity in the 2--16 keV 
band (the intensity is in Crab units). We believe that this QPO was always 
the lower kHz QPO because in our data this was always the strongest one 
(a comparison between the rms amplitudes of both kHz QPOs is
in Fig.~\ref{fig:fig3}, top panel). 
At least two parallel tracks are visible in this figure, demonstrating 
that there is a complex long-term relation between kHz QPO frequencies and 
source count rate.  Different colors correspond to different dataset;
P10088 (from 1996 April 27 to 1996 April 30, corresponding to the red 
points at the low-flux level), P10072 (data taken in 1996 May 29, 
black points); P10088 (from 1996 November 9 to 1997 January 6, corresponding 
to the red points at the high-flux level); P30053 (1998 February 24--25 
and 1998 August 19--20, gray points).
On long timescales, the source does not move randomly on the kHz frequency
versus flux diagram, but jumps back and forth between the same two tracks.
Although the observations are not always continuous, we can see that 
at the beginning of our observing period in April 1996 the source is found 
at the low flux level, where it remains till the end of May 1996, 
then it jumps to the high-flux track where it stays from
1996 November 9 to 1997 January 6, and finally the source comes back to the
low-flux track where it is found in 1998 February 24--25 and 1998 August 
19--20.  The difference in intensity between these two tracks is 
$\sim 20\%$, similar to the intensity shift observed in the hardness-intensity 
diagram.  Indeed the tracks at low and high flux levels correspond, 
respectively, to the data points at the left-hand ends of the 
hardness-intensity diagram, where the same $\sim 20\%$ intensity shift 
is observed.  

These two parallel tracks almost overlap each other 
when we plot the lower peak frequency against the colors 
(Fig.~\ref{fig:fig2}, bottom panels) or $S_{\rm a}$ (Fig.~\ref{fig:fig2}, 
right top panel).  However, significant shifts between different
observations are still visible in these plots. 
In particular, datasets corresponding to lower flux levels are found towards 
the left (closer to the island state) in the color-color diagram 
(i.e.\ at lower $S_{\rm a}$ values), while data corresponding to higher flux 
levels have slightly higher $S_{\rm a}$ values.
Similar systematic shifts are also observed in the plots of the kHz QPO 
frequency versus hard and soft colors, with the source having higher hard 
colors and lower soft colors at low flux levels, and vice versa 
(see Fig.~\ref{fig:fig2}). 
The shifts in the QPO frequencies vs.\ $S_{\rm a}$ diagram are almost certainly 
due to the fact that we consider a single spline across the color-color 
diagram, while there are two different atoll tracks corresponding to the two 
different flux levels.
Note however that there is an intrinsic spread in the QPO frequency vs.\ 
soft color and vs.\ $S_{\rm a}$ relations within the dataset P30053 (gray 
points), although these all show the same flux level and seem to belong to 
the same atoll track.

Although the data show some scatter, the frequency of the kHz QPO seems 
to be slightly anticorrelated with the hard color; fitting a line to the 
hard color vs.\ frequency relation significantly improves the fit with 
respect to a constant function; the slope is different from zero at the
$\sim 7 \sigma$ level for the dataset P10088 (red points), and at 
the $\sim 4 \sigma$ level for the dataset P30053 (gray points). 
The kHz QPO frequency is slightly correlated with the intensity, 
whereas there is no significant correlation between 
the kHz QPO frequency and the soft color or $S_{\rm a}$. 

On several occasions we also detected the upper kHz QPO, either
directly in the dynamical power spectrum, or in the average 
power spectrum of an observation.
To increase the significance of this QPO and to facilitate measuring its
parameters, we needed to average several 64-s power spectra.
Because there seems to be no clear correlation between the 
frequency of the kHz QPOs and $S_{\rm a}$, 
we used the frequency of the lower kHz QPO   
to group the power spectra.  Following the procedure described in 
M\'endez et al.\ (1998b), we aligned the individual power spectra according 
to the frequency of the lower peak, we collected them into groups such 
that the frequency of this QPO did not vary by more than $15-20$ Hz, and 
we calculated the average power spectrum for each group.
Because we want now to measure the rms amplitudes of the QPOs,
here we only used data for which the full energy band was available; 
based on this requirement we discarded observation P10072, corresponding 
to $1.2\%$ of the data. (Note that this restriction was not necessary 
before because then we only measured the QPO frequencies).
The measured frequencies, rms amplitudes, and quality factors of both kHz 
QPOs are plotted in Fig.~\ref{fig:fig3} as a function of the lower peak 
frequency.  The frequency of the lower peak ranges between 780 and 910 Hz. 
The rms amplitude of the lower peak is approximately constant (or slowly 
decreasing) up to $\nu_{\rm low} \sim 850$ Hz,
and then gradually decreases from $\sim 7\%$ to $\sim 4\%$ while 
$\nu_{\rm low}$ increases from $\sim 850$ Hz to $\sim 910$ Hz. The 
upper peak was significantly detected only for $\nu_{\rm low} \ga 840$~Hz.  
Its frequency increases from $\sim 1100$ to $\sim 1150$
Hz when the frequency of the lower peak increases from $\sim 840$ Hz to 
$\sim 910$ Hz, while its rms amplitude remains constant around $2.5-3\%$.
While the quality factor of the upper kHz QPO is compatible with being
constant at a value of $15-20$, the quality factor of the lower kHz QPO
significantly decreases from $\sim 200$ to $\sim 50$ with increasing 
frequency, implying that the lower peak becomes broad at high frequencies.
In those intervals in which the upper kHz QPO was not detected with high
statistical significance, we calculated the $90\%$ confidence level upper 
limits on its rms amplitudes, which are still $\sim 3-4\%$ (assuming a 
frequency separation between the two kHz QPOs of 260 Hz, see below, and a 
typical quality factor for the upper kHz QPO of 16).

\begin{figure}
\hbox{\hspace{0cm}\psfig{figure=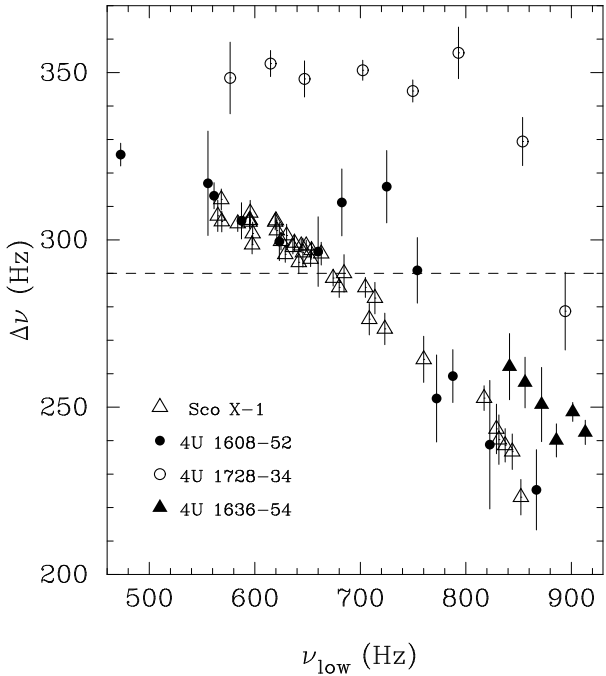,width=8.0cm}}
\caption[]{The kHz QPO frequency separation as a 
function of the frequency of the lower peak for Sco X--1
(open triangles, van der Klis et al.\ 1997), 4U 1608--52 (filled circles, 
M\'endez et al.\ 1998c), 4U 1728--34 (open circles, M\'endez \& van der Klis
1999), and \1636 (filled triangles, this work). Half the frequency of 
the burst oscillations, $\nu_{\rm burst}/2 \simeq 290$ Hz, for \1636 is also 
shown (dashed line, e.g.\ Strohmayer et al.\ 1998).}
\label{fig:fig3b}
\end{figure}

The difference between the frequencies of the kHz QPOs generally decreases
from $262.1 \pm 9.9$ Hz to $242.4 \pm 3.6$ Hz, although the decrease, when 
compared with the associated uncertainties, is significant only at a $1.8 
\sigma$ level (an F-test yields a probability in favor of the hypothesis 
that the frequency difference decreases of $\sim 75\%$).  
Although the decrease is not highly statistical significant,
the trend shown by \1636 is in agreement with the trend shown
by other sources (see Fig.~\ref{fig:fig3b}). Also, the frequency separation 
between the two kHz QPOs is significantly smaller than half the frequency of 
the quasi-coherent oscillations (the dashed line in Fig.~\ref{fig:fig3b}) 
that have been observed in this source during type-I X-ray bursts, in 
agreement with the results of M\'endez et al.\ (1998a).

\subsection{Broad band power spectra}

To study the time variability of the source over a wide frequency range
as a function of the position in the color-color diagram we divided this 
diagram into several intervals. As already noted in the previous section,
some shifts are present in the color-color diagram of all the data combined.
In principle we could try to correct for this by drawing two splines,
one for each of the two parallel tracks, as it is usually done in the case of 
Z sources to correct for the secular shifts of the Z track (see e.g.\ 
Kuulkers et al.\ 1994; Jonker et al.\ 1998; Jonker et al.\ 2002b).
However, in the case of \1636 (and of atoll sources in general) this method
cannot be applied, because there are no precise vertices in the atoll track, 
which are instead quite well defined in a Z track. This means that if we draw
two splines for the two parallel atoll tracks, the S parametrization along 
these would be arbitrary, and it would be impossible to compare the $S_{\rm a}$ 
values corresponding to one track with the other. Another possibility is 
to use the QPO frequency vs.\ $S_{\rm a}$ relation to correct for these shifts.
As in \1636 QPOs are only detected in a small part 
of the color-color diagram (a small range of $S_{\rm a}$) we cannot use 
these to correct for the shifts in the color-color diagram or to select the 
power spectra on the basis of the QPO frequencies. 
In the following, considering that the shifts in the color-color diagram are 
much smaller than in the hardness-intensity diagram, we select the power 
spectra according to their position in the color-color diagram of all the 
data combined, keeping in mind that they might not correspond exactly to the 
same position relatively to the atoll track. 
When possible we analyze separately power spectra 
corresponding to the same $S_{\rm a}$ value but to different flux levels  
to see if there are differences between them.

\begin{table*}
\caption[2a]{Parameters of the very low frequency noise and the band limited 
noise components.
}
\begin{center}
\tiny
\label{tab2}
\vskip 0.2cm
\begin{tabular}{ccccccccccc}
\hline  
         &    &   & \multicolumn{2}{c}{VLFN} &   & \multicolumn{3}{c}{BLN}  &  \\ 
\cline{4-5} \cline{7-9} \\
Int. & $S_{\rm a}$ & Rate$^{\mathrm{a}}$ & $\alpha$ & rms$^{\mathrm{b}}$ &    &
$\nu_{\rm max}$ & $Q$  &rms & $\chi^2/d.o.f.$ & F-test$^{\mathrm{c}}$ \\ 
Num.   &       & (c/s) &       & (\%)&     &  (Hz)     &   & (\%)  &   & \\ 
\hline
1 & 0.75--1.1 & 1793.59 & $1.386 \pm 0.071$ & $2.01 \pm 0.27$ &   & 
$69 \pm 12$ & $0.20 \pm 0.15$ & $3.70 \pm 0.23$ & 309/136 & $1.5 \times 10^{-3}$ \\
2 & 1.1--1.25 & 2016.86 & $1.454 \pm 0.027$ & $2.458 \pm 0.071$ &   &
$59.7 \pm 3.8$ & $0.362 \pm 0.072$ & $3.27 \pm 0.10$ & 358/137 & $2.4 \times 10^{-3}$ \\
3 & 1.25--1.4 & 2215.07 & $1.413 \pm 0.020$ & $2.742 \pm 0.074$ &   &
$56.2 \pm 3.1$ & $0.494 \pm 0.078$ & $2.994 \pm 0.093$ & 153/137 & $1.5 \times 10^{-6}$ \\
4 & 1.4--1.55 & 2431.22 & $1.372 \pm 0.046$ & $2.81 \pm 0.33$ &   &
$57 \pm 11$ & 0 (fixed) & $2.70 \pm 0.17$ & 129/137 & $2.9 \times 10^{-8}$ \\
5 & 1.55--1.7 & 2660.44 & $1.371 \pm 0.087$ & $3.62 \pm 0.21$ &   &
$79 \pm 16$ & 0 (fixed) & $2.59 \pm 0.16$ & 146/137 & $3.3 \times 10^{-8}$ \\
6 & 1.7--1.85 & 2941.12 & $1.327 \pm 0.021$ & $3.81 \pm 0.13$ &   & 
$109 \pm 29$ & 0 (fixed) & $2.41 \pm 0.20$ & 162/138 & $1.0 \times 10^{-4}$ \\
7 & 1.85--2 & 3148.64 & $1.345 \pm 0.023$ & $3.80 \pm 0.14$ &   &
$46 \pm 15$ & 0 (fixed) & $2.05 \pm 0.18$ & 172/138 & $2 \times 10^{-8}$ \\
8-low & 2--2.2 & 2626.62 & $1.459 \pm 0.050$ & $4.01 \pm 0.24$ &   &
$17.9 \pm 8.6$ & 0 (fixed) & $1.72 \pm 0.33$ & 138/140 & --  \\
8-high& 2--2.2 & 3382.33 & $1.430 \pm 0.038$ & $3.66 \pm 0.23$ &   &
$34.1 \pm 9.2$ & 0 (fixed) & $2.04 \pm 0.37$ & 133/137 & $4.2 \times 10^{-4}$ \\
9-low & 2.2--2.45 & 2693.08 & $1.608 \pm 0.061$ & $4.68 \pm 0.37$ &   &
$20.7 \pm 7.7$ & 0 (fixed) & $2.41 \pm 0.29$ & 162/140 & --  \\
9-high& 2.2--2.45 & 3543.68 & $1.422 \pm 0.048$ & $3.47 \pm 0.25$ &   &
$24 \pm 13$ & 0 (fixed) & $1.59 \pm 0.27$ & 123/138 & $2.7 \times 10^{-3}$ \\
10 & 2.45--3 & 3332.48 & $1.58 \pm 0.11$ & $4.21 \pm 0.79$ &   &
$22 \pm 12$ & $0.31 \pm 0.54$ & $2.0 \pm 0.43$ & 62/92 & --  \\
\hline
\end{tabular}
\end{center}
Errors correspond to $\Delta \chi^2 = 1$.
\begin{list}{}{}
\item[$^{\mathrm{a}}$] PCA count rate, 5 PCUs, not corrected for the background.
The background average count rate of the PCA is 132 c/s.
\item[$^{\mathrm{b}}$] rms amplitude calculated in the frequency range 
0.001--1.0 Hz. 
\item[$^{\mathrm{c}}$] F-test is the probability of chance improvement of the
fit when the very low frequency Lorentzian is included.
\end{list}
\end{table*}

\begin{table*}
\caption[2b]{Parameters of the very low frequency Lorentzian and the kHz QPOs.}
\begin{center}
\tiny
\label{tab2b}
\begin{tabular}{cccccccccccc}
\hline  
     & \multicolumn{3}{c}{VLF Lor} &  & \multicolumn{3}{c}{Lower kHz QPO} & 
     & \multicolumn{3}{c}{Upper kHz QPO} \\ 
\cline{2-4} \cline{6-8} \cline{10-12} \\
Int. & $\nu_{\rm max}$ & $Q$  &rms & & $\nu_{\rm max}$ & $Q$  
&rms& & $\nu_{\rm max}$ & $Q$  & rms \\ 
Num.   &  (mHz)     &   & (\%) &  & (Hz)  &   & (\%) & &  (Hz)  &   & (\%) \\ 
\hline
1 & $25.5 \pm 0.38$ & $0.40 \pm 0.31$ & $1.01 \pm 0.29$ & &
$864.6 \pm 1.2$ & $12.49 \pm 0.63$ & $6.50 \pm 0.12$ & &
$1152.8 \pm 6.6$ & $19.8 \pm 6.8$ & $2.67 \pm 0.28$ \\
2 & $28.6 \pm 4.6$ & 0.8 (fixed) & $0.597 \pm 0.058$ & &
$866.2 \pm 1.4$ & $10.23 \pm 0.37$ & $5.646 \pm 0.077$ & &
$1161.2 \pm 4.4$ & $15.8 \pm 2.3$ & $2.71 \pm 0.15$ \\
3 & $25.0 \pm 2.6$ & 0.8 (fixed) & $0.685 \pm 0.067$ & &
$901.7 \pm 1.5$ & $14.6 \pm 1.2$ & $3.69 \pm 0.10$ & &
$1193.0 \pm 3.6$ & $15.3 \pm 2.1$ & $2.94 \pm 0.14$ \\
4 & $23.7 \pm 9.5$ & $< 0.27$ & $1.40 \pm 0.39$ & &
$911 \pm 45$ & $3.2 \pm 1.3$ & $2.92 \pm 0.38$ & &
$1227.7 \pm 2.7$ & $52 \pm 14$ & $1.69 \pm 0.17$ \\
5 & $27.8 \pm 9.2$ & $0.58 \pm 0.47$ & $1.2^{+1.4}_{-0.24}$ & &
-- & -- & -- & & 
$1233.1 \pm 9.0$ & $25 \pm 15$ & $1.31 \pm 0.26$ \\
6 & $19.2 \pm 2.0$  & 0.8 (fixed) & $1.16 \pm 0.13$ & &
-- & -- & -- & & 
-- & -- & --  \\
7 & $23.7 \pm 1.8$  & 0.8 (fixed) & $1.41 \pm 0.11$ & &
-- & -- & -- & & 
-- & -- & --  \\
8-low  & 25 (fixed) & 0.8 (fixed) & $< 0.54$ & & 
-- & -- & -- & & 
-- & -- & --  \\
8-high & $27.1 \pm 2.9$ & 0.8 (fixed) & $1.13 \pm 0.16$ & &
-- & -- & -- & & 
-- & -- & --  \\
9-low  & 25 (fixed) & 0.8 (fixed) & $< 0.63$ & &
-- & -- & -- & & 
-- & -- & --  \\
9-high & $29.4 \pm 4.9$ & 0.8 (fixed) & $1.06 \pm 0.18$ & &
-- & -- & -- & & 
-- & -- & --  \\
10 & 25 (fixed)  & 0.8 (fixed) & $1.32 \pm 0.42$ & & 
-- & -- & -- & & 
-- & -- & --  \\
\hline 
\end{tabular}
\end{center}
Errors correspond to $\Delta \chi^2 = 1$, while upper limits to the rms 
amplitude of the very low frequency Lorentzian correspond to $90\%$ 
confidence level.
\end{table*}

We divided the color-color diagram of \1636 into 10 intervals corresponding 
to different consecutive values of $S_{\rm a}$.  
We defined the intervals such that we had a sufficient number of them, and 
enough data in each of them to have good statistics. Each interval contains 
between 15 and 370 power spectra.  For each interval
we calculated the average $S_{\rm a}$, using the same spline parametrization 
described in Sect. 3; the intervals are indicated with numbers from 1 to 10 in 
order of increasing $S_{\rm a}$.
For each interval we computed a power spectrum dividing
the PCA light curve into 256-s long segments for which we calculated
Fourier power spectra, over the whole PCA energy range, up to a Nyquist
frequency of 2048 Hz. We averaged together all power spectra
that fell in one interval for more than 40\% of the time, weighting each
of them by the fraction of the time the source actually spent in the interval. 
We produced two different power spectra each for intervals 
8 and 9, one at a low and one at a high flux level, respectively. 
For interval 10 the statistics were not enough to produce two separate 
power spectra, so we preferred to average them together.
On average the count rate increases with increasing $S_{\rm a}$ 
(see Table~\ref{tab2}), except for the spectra from intervals 8 and 9 
corresponding to lower flux, and for interval 10 (note that in this interval 
the low and high flux power spectra are averaged together).
The high frequency part of each power spectrum ($1200-2048$ Hz, where neither 
noise nor QPOs are known to be present) was used to estimate the Poisson 
noise, which was subtracted before fitting the power spectra.

\begin{figure*}
\hbox{\hspace{0cm}\psfig{figure=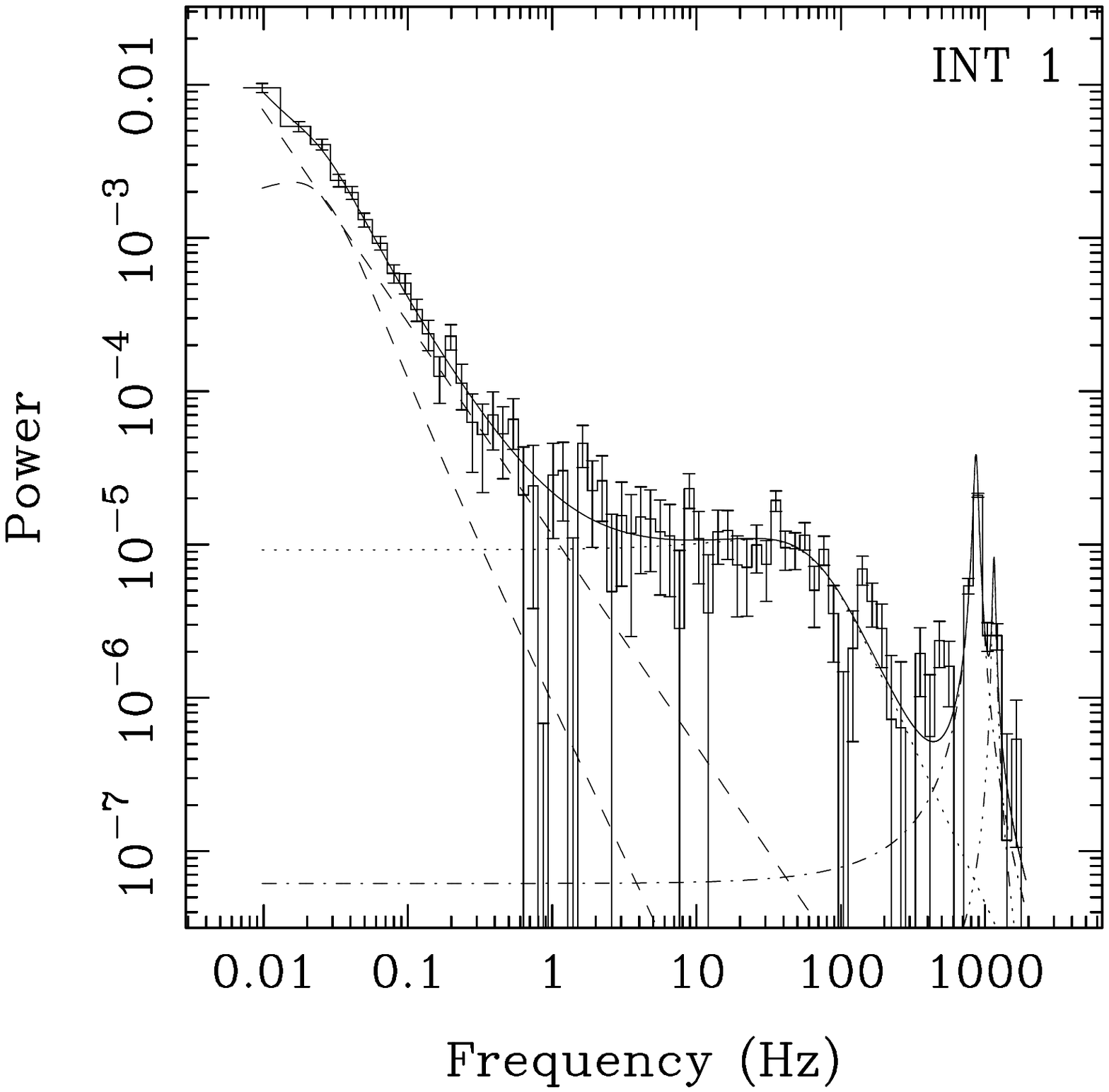,width=8.0cm}\hspace{1cm}
\psfig{figure=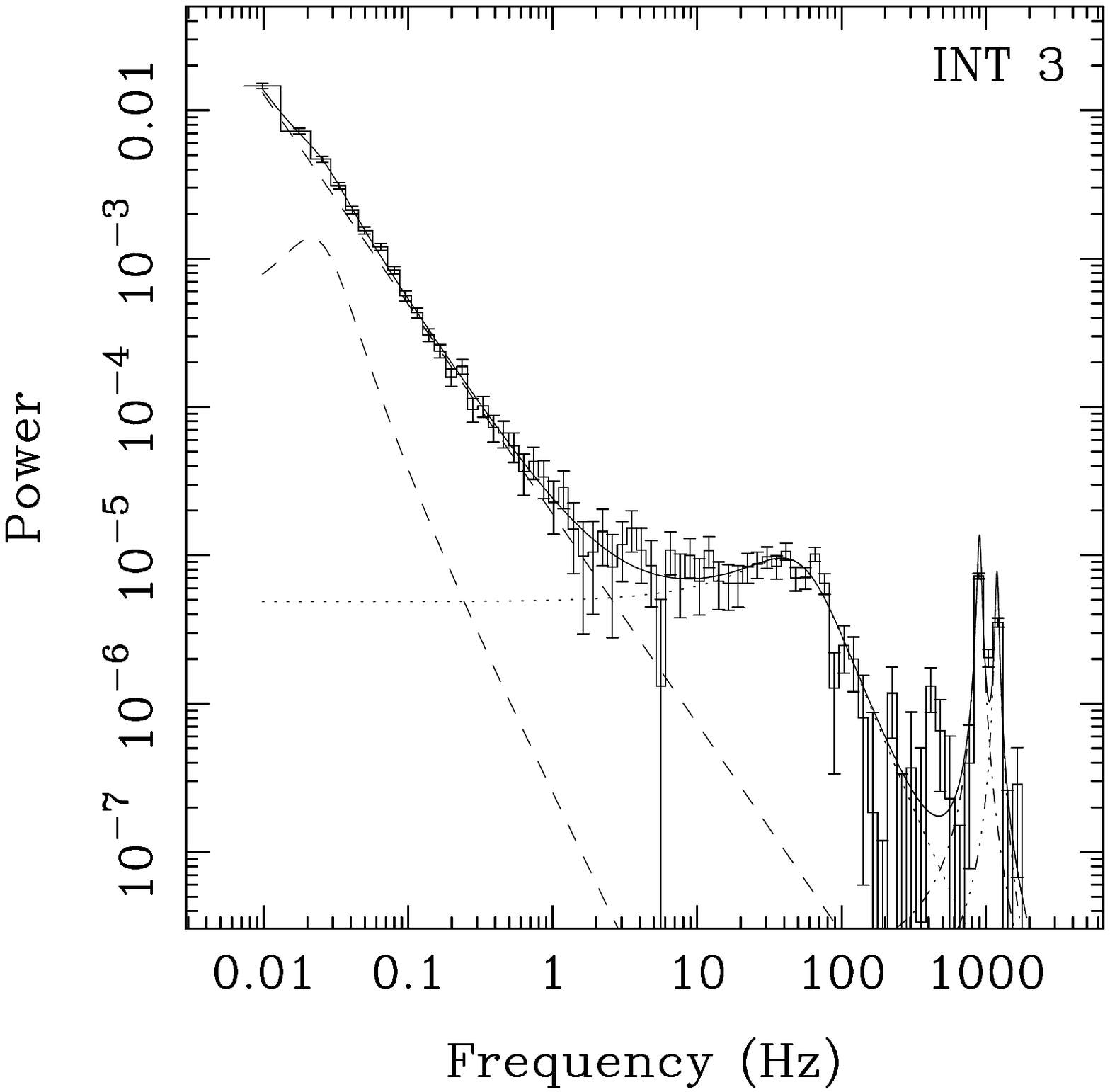,width=8.0cm}}
\hbox{\hspace{0cm}\psfig{figure=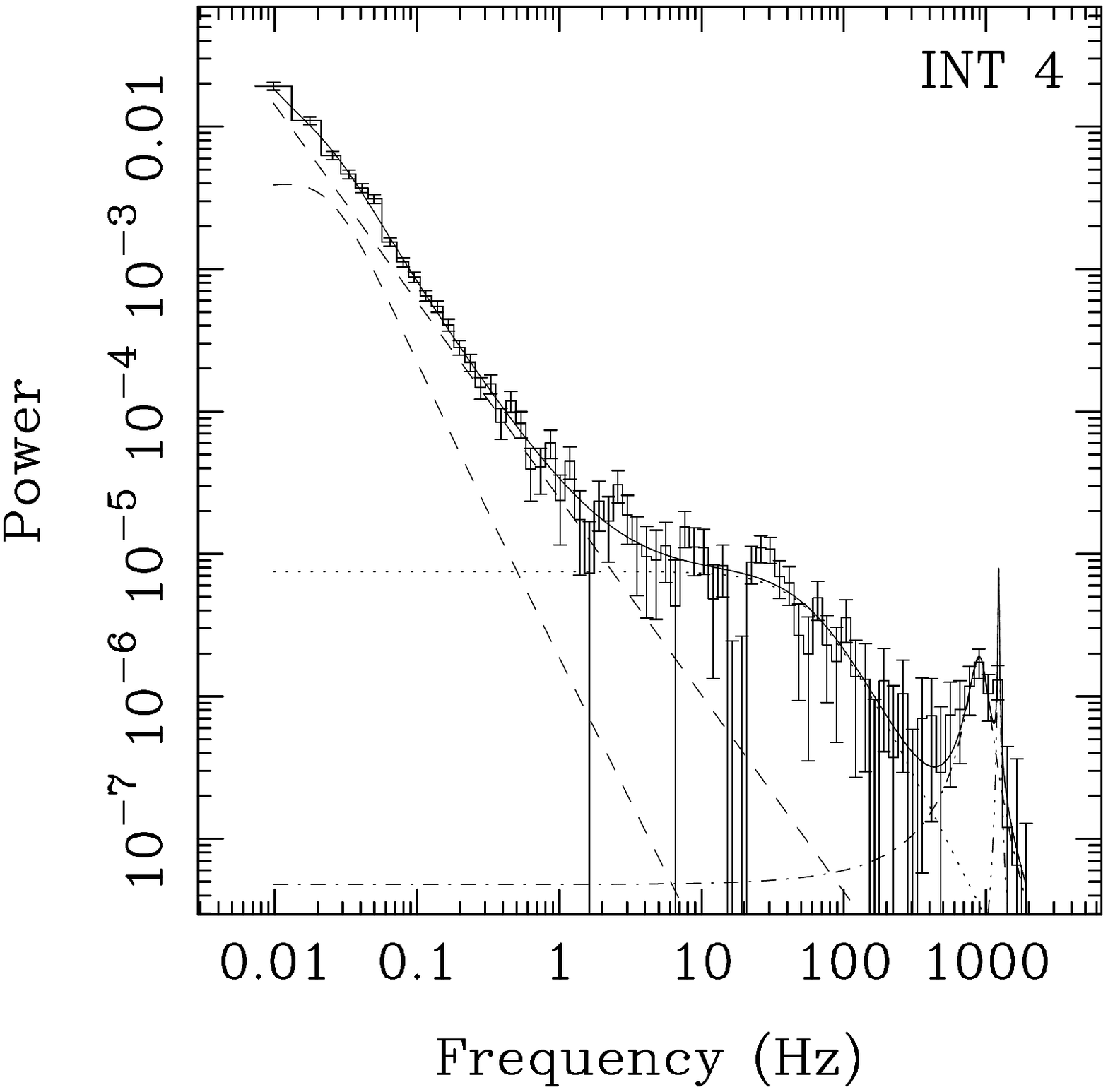,width=8.0cm}\hspace{1cm}
\psfig{figure=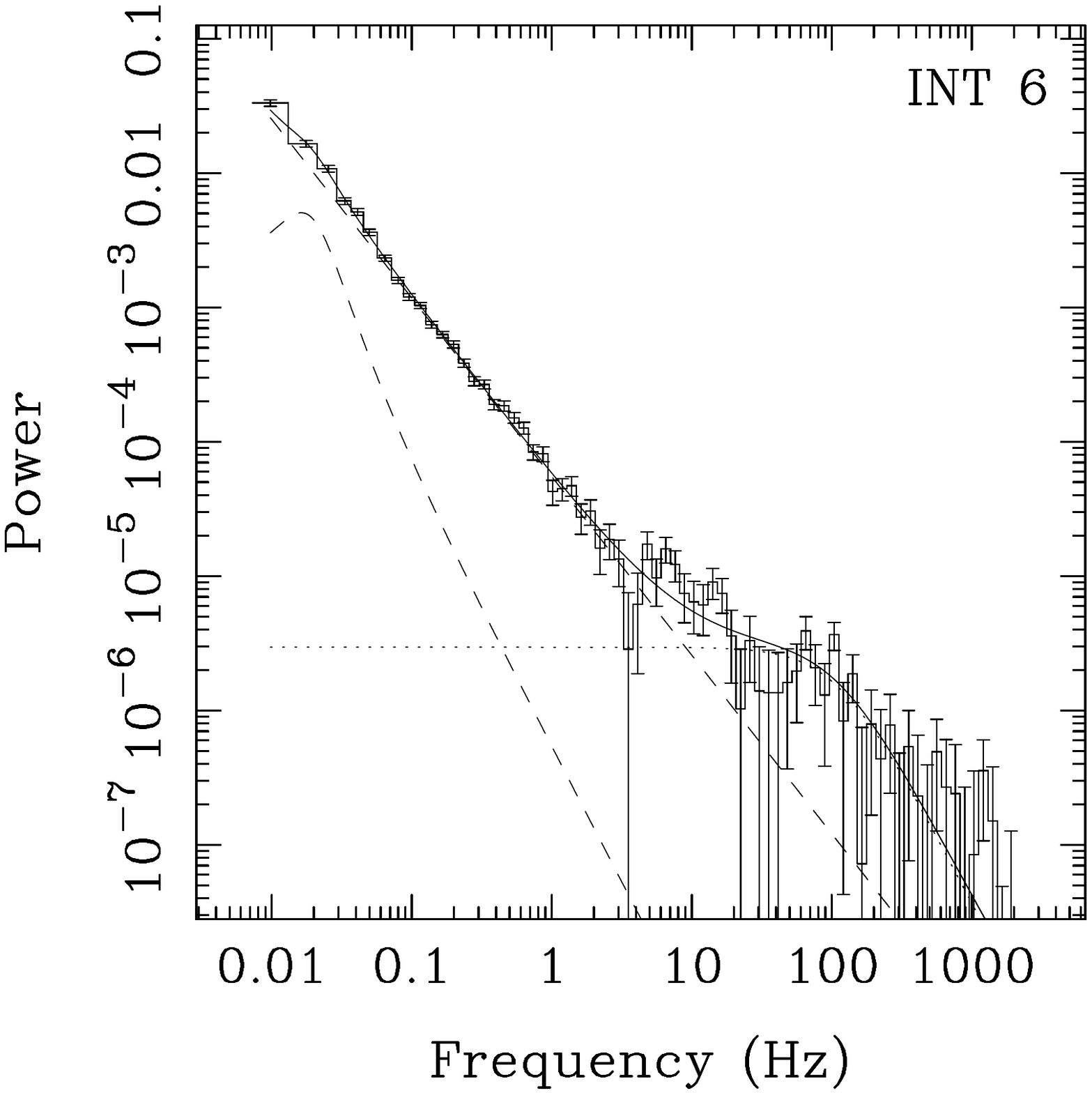,width=8.0cm}}
\hfill      \parbox[b]{18cm}{\caption[]{Representative power spectra of \1636 
corresponding to different intervals in the color-color diagram of Fig.~\ref{fig:fig1}; 
the power units are (rms/mean)$^2$/Hz. 
Interval numbers are indicated in the top right corners. 
For each interval, the power spectrum and the best-fit model components are 
shown; from lower to higher frequencies these components are the very low 
frequency noise (dashed line), the very low frequency Lorentzian (dashed 
line), the band limited noise (dotted line), and the 
kHz QPOs (dot-dashed line and dot-dot-dot-dashed line, respectively, 
when significantly detected). Note that the parameters for the 
kHz QPOs were determined using a higher frequency resolution than that
used for the broad band power spectra shown here.
}\label{fig:fig3c}}
\end{figure*}

In Fig.~\ref{fig:fig3c} we show some representative power spectra 
(rms-normalized and Poisson-noise subtracted) corresponding to intervals 
1, 3, 4, and 6.
All of them are typical of the banana state, showing the VLFN, dominating
up to frequencies of the order of 0.1 Hz, and a BLN component, dominating
up to frequencies of the order of 100 Hz. The power spectra of intervals 
1 to 5 also show one or two kHz QPOs.
We used a power law, $P(\nu) \propto \nu^{-\alpha}$, to fit the VLFN
and Lorentzians to fit the BLN and the kHz QPOs, when present. 
We fitted the kHz QPOs, when significantly detected, in a limited frequency 
range (500--2048 Hz) where no other component was present, and using a higher
frequency resolution.  We then fitted the other components using the full 
frequency range and fixing the parameters of the kHz QPOs at the values 
obtained in the high frequency range.  This could be done because (as we 
verified) the frequencies of the kHz QPOs were high enough that they do not 
affect the parameters of the low frequency features and vice versa.

The VLFN and the BLN components were present in all the power spectra,
while the two kHz QPOs were significantly detected (at more than 
$3 \sigma$) only in the first four intervals.  In interval 5 only one
kHz QPO is significantly detected, probably the upper kHz QPO given its
high frequency ($\sim 1230$ Hz).  For some of the intervals these components 
are not sufficient to give a good fit to the broad band power spectrum.  
In those cases, some residuals are present between 0.01 and 0.1 Hz
due to the VLFN not being well represented by a power law.
This is specially apparent in intervals 3 to 5 and 7.   
We therefore added another Lorentzian to the model, with characteristic 
frequency $\nu_{\rm max}$ between 20 and 30 mHz. 
This Lorentzian significantly improved the fit in these intervals. 
This feature is unlikely to be due to the averaging process, i.e.\
to the fact that we combine power spectra that may have slightly 
different characteristic frequencies; in fact we still find it, although 
at a lower significance level if we divide interval 4 in three 
subintervals (containing $\sim 10$ power spectra each); sometimes 
this component is detected in the power spectrum of a single observation. 
For interval 4 we also tried to fit the complex shape of the VLFN 
using a broken power law (with a break frequency around 20 mHz) instead of 
a power law plus a Lorentzian. This model gives a fit of similar quality 
($\chi^2 = 133$ for 139 dof) with respect to the power law plus Lorentzian
model.
In some of the other intervals the addition of this Lorentzian at mHz 
frequencies was still statistically significant. However, sometimes we 
could not obtain a stable fit keeping all the parameters free. 
In those cases, because the goodness of the fit was less sensitive to 
the quality factor of this Lorentzian than to the other parameters, 
we fixed $Q$ to a value of 0.8 and let the frequency and the rms 
amplitude be free. For intervals 8 low-flux, 9 low-flux and 10, the 
fit was not improved by the addition of this component and
we could only find an upper limit to its rms amplitude.

The results of these fits are shown in Tables~\ref{tab2} and \ref{tab2b}. 
The parameters of the Lorentzian components are expressed in terms of the 
integrated fractional rms amplitude, the quality factor $Q$ ($Q \equiv 
\nu_0/2\Delta$, where $\nu_0$ and $\Delta$ are the centroid 
frequency and the HWHM of the Lorentzian, respectively), 
and the frequency $\nu_{\rm max} = \nu_0 \sqrt{1 + 1/(4 Q^2)} = 
\sqrt{\nu_0^2 + \Delta^2}$ at which the Lorentzian function has its maximum 
in a $\nu P (\nu)$ representation (Belloni et al.\ 2002).  
For narrow features ($Q > 1$) $\nu_{\rm max}$ is very close to the centroid 
frequency $\nu_0$, while for Lorentzians with $Q \la 1$, $\nu_{\rm max}$ 
approaches its HWHM, $\Delta$. We obtained good fits for all the intervals 
with reduced $\chi^2$ ranging from 0.7 to 2.5 for 92 to 140 dof. 
Some of the $\chi^2$ values are rather large, but
this is not surprising given the high statistics of the power spectra. 
Still, for all the intervals the residuals are usually within 2 
$\sigma$ and featureless. The largest $\chi^2$ values occur for the 
first two intervals. A careful examination of the residuals shows that the 
largest contributions to the $\chi^2$ come from the region above 500 Hz,
and are due to the fact that we fitted the parameters of the kHz QPOs 
using a higher frequency resolution (these parameters are not therefore
the best fit parameters in the broad band and more rebinned power spectra). 
Note that while the best fit parameters obtained for intervals 8 low-flux and 
8 high-flux are compatible with each other (except for the presence of the 
very low frequency Lorentzian), some (marginal) differences seem to be present 
between intervals 9 low-flux and 9 high-flux. These differences will be noted 
below, if necessary. 
In the following we will describe the results in more detail.


\begin{figure}
\hbox{\hspace{0cm}\psfig{figure=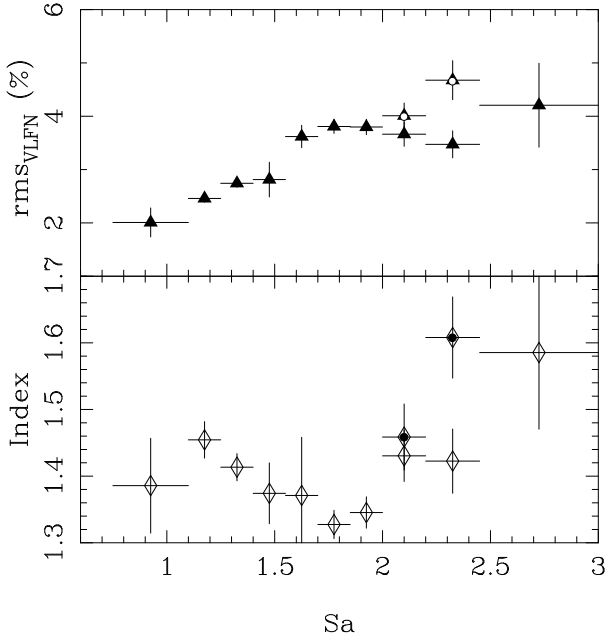,width=8.0cm}}
\caption[]{Very low frequency noise (power-law component) rms amplitude (top 
panel) and power-law index (bottom panel) as a function of $S_{\rm a}$ for 
\1636.  The points marked with a white or black dot correspond to the 
intervals at low flux.}
\label{fig:fig4} 
\end{figure}
\begin{figure}
\hbox{\hspace{0cm}\psfig{figure=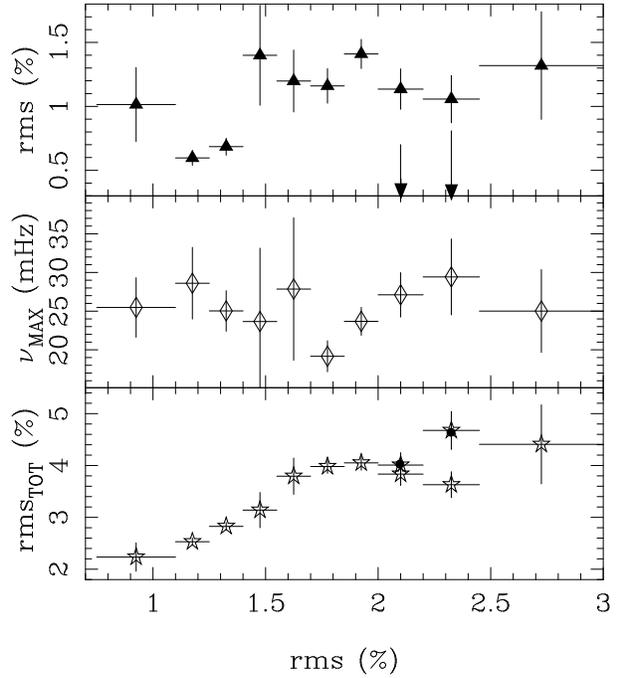,width=8.0cm}}
\caption[]{Very low frequency Lorentzian rms amplitude (top panel) and 
characteristic frequency (middle panel) as a function of $S_{\rm a}$.
The arrows in the top panel indicate the $90\%$ confidence level
upper limits calculated for the intervals at low flux.
In the bottom panel the total very low frequency rms amplitude (the combined 
very low frequency Lorentzian and very low frequency noise rms amplitude) 
is plotted versus $S_{\rm a}$.
The points marked with a black dot correspond to the intervals at low flux.}
\label{fig:fig4b} 
\end{figure}

In Fig.~\ref{fig:fig4} we plot the rms amplitude and power-law index of the 
VLFN as a function of $S_{\rm a}$. While no clear trend is visible in the 
power-law index, which oscillates around a value of $\sim 1.5$ (probably 
slightly increasing with $S_{\rm a}$), the VLFN rms amplitude first increases 
from $2\%$ to $\sim 4\%$ at low values of $S_{\rm a}$, and then saturates at 
$\sim 4\%$ for $S_{\rm a}$ higher than 1.5. Both the fractional rms amplitude 
and the power-law index of the VLFN are higher in interval 9 low-flux  
than in 9 high-flux (the differences are $\sim 2.7 \sigma$ and $\sim 2.4 
\sigma$, respectively). 
This difference might be due to the presence of an extra component, 
i.e.\ the very low frequency Lorentzian, in interval 9 high-flux, which
absorbs part of the power. However, looking at the combined rms amplitude
of the VLFN and very low frequency Lorentzian (Fig.~\ref{fig:fig4b}, bottom
panel), we can see that the total (very low frequency) rms amplitude of 
interval 9 low-flux is still higher (at $\sim 2.4 \sigma$ confidence level)
than that of interval 9 high-flux.
On the other hand, no significant differences are visible between intervals
8 low-flux and 8 high-flux.
In 4 out of the 10 intervals we significantly detected an excess of power 
between 0.01 and 0.1 Hz with respect to the power law used to fit the VLFN. 
We fitted this excess with a (broad) Lorentzian, the parameters of which are 
shown in Fig.~\ref{fig:fig4b} (top and middle panels).  
The rms amplitude of this feature is between 1\% and 1.5\% (except
for intervals 2 and 3 where it is $\sim 0.6\%$), the quality factor $Q$ is 
rather low (less than 0.8), and the characteristic frequency, 
$\nu_{\rm max}$, is between 20 and 30 mHz.
The addition of this feature did not significantly improve the fit in 
intervals 8 low-flux, 9 low-flux and 10. In particular, in intervals 8 low-flux
and 9 low-flux, the $90\%$ confidence level upper limits 
on the very low frequency Lorentzian rms amplitude are $\sim 0.7-0.8\%$, which 
is significantly lower than the rms amplitudes measured for most of the 
other intervals.

\begin{figure}
\hbox{\hspace{0cm}\psfig{figure=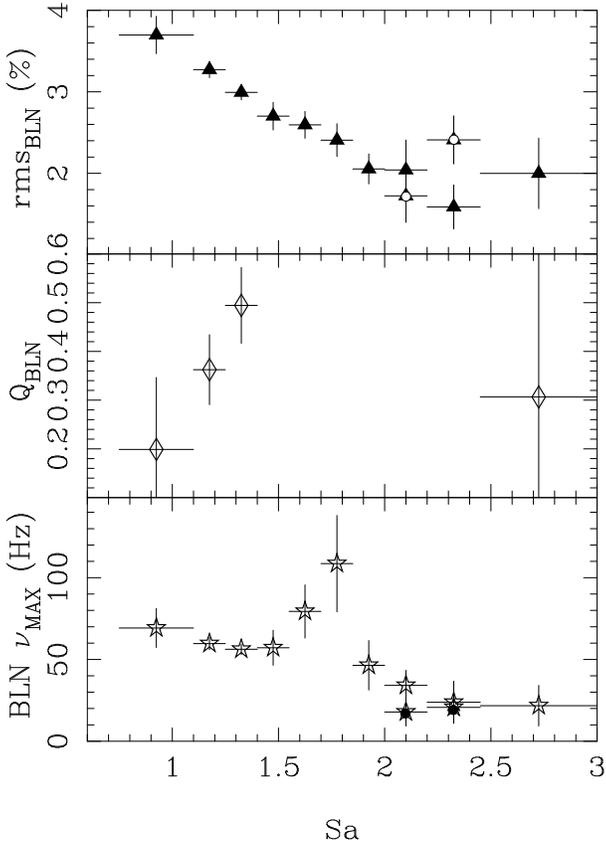,width=8.0cm}}
\caption[]{Rms amplitude (top panel), quality factor 
(middle panel), and characteristic frequency $\nu_{\rm max}$ (bottom panel)
of the band limited noise component in \1636. The points marked with a white 
or black dot correspond to the intervals at low flux.}
\label{fig:fig5}
\end{figure}

The parameters of the BLN are plotted versus $S_{\rm a}$ in 
Fig.~\ref{fig:fig5}. 
The rms amplitude (Fig.~\ref{fig:fig5}, top panel) decreases from 4\% to 
$\sim 2\%$ with increasing $S_{\rm a}$ up to $S_{\rm a} = 2$, and above that 
value it seems to remain constant at $\sim 2\%$.
Again, the rms amplitude for interval 9 low-flux is slightly higher 
($\sim 2.1 \sigma$) than that for 9 high-flux.
The quality factor $Q$ of the BLN (Fig.~\ref{fig:fig5}, middle panel) is always 
less than $\sim 0.5$. Most of the times (intervals 4 to 9) we fixed it to 0
because it could not be determined from the fit, which gave negative values
but still compatible with 0. The characteristic frequency $\nu_{\rm max}$ 
of the BLN (Fig.~\ref{fig:fig5}, bottom panel) tends 
to decrease from $\sim 70$ Hz to $\sim 20$ Hz with increasing $S_{\rm a}$; 
a linear fit of these points gives a slope of $-24.4 \pm 5.8$ Hz per unit 
$S_{\rm a}$, with a probability of chance improvement with respect to a fit 
with a constant of $\sim 3.6 \times 10^{-3}$. 
However, in intervals 5 and 6, $\nu_{\rm max}$ seems to increase up to 
$\sim 100$ Hz (although the uncertainties on these two points are quite 
large; these points deviate from the linear fit by $\sim 2\sigma$).

From the broad band spectra we have also measured the parameters of the kHz 
QPOs as a function of $S_{\rm a}$ for those intervals (from 1 to 5) in which 
one or two kHz QPOs were detected at more than $3 \sigma$ (see 
Table~\ref{tab2b}).  
The frequency of both the kHz QPOs increases with $S_{\rm a}$, as expected.
The rms amplitude of the lower kHz QPO decreases from $\sim 6\%$ to $\sim 3\%$ 
with increasing $S_{\rm a}$. The rms amplitude of the upper kHz QPO slightly 
increases in intervals 1 to 3 ($S_{\rm a} < 1.4$, $\nu_{\rm low} \la 900$ Hz), 
and then significantly decreases from $\sim 3\%$ to $\sim 1\%$ at higher 
$S_{\rm a}$ (intervals 4 and 5).
The small differences with respect to the kHz QPOs rms amplitudes shown in
Fig.~\ref{fig:fig3} are due to the different way in which the power spectra
were grouped and averaged together in the two cases. Using the $S_{\rm a}$ 
values to group the power spectra, we could average together more power 
spectra; this allows us to see the upper peak up to higher frequencies, but 
also introduces an artificial broadening of the QPOs due to the fact that 
their frequencies change with $S_{\rm a}$.

\section{Discussion}

We have analyzed $\sim 600$ ks of RXTE data of \1636 taken between
April 1996 and February 1999. We studied the
correlated spectral and timing properties of this source using broad 
band (4 mHz to 2048 Hz) power spectra extracted at different
positions of the source in the color-color diagram. 
The most notable results of this work are: a) The presence of shifts,
due to secular motion of the atoll track, in both the hardness-intensity 
and color-color diagrams; b) The presence of parallel tracks in the 
kHz QPO frequency versus intensity diagram; for the first time we see 
that these tracks do not overlap perfectly when plotted versus colors or 
$S_{\rm a}$; this is probably a consequence of the secular motion of the atoll 
track mentioned above; c) The deviation of the VLFN-component shape from 
a pure power law in some of the intervals; d) The characteristic frequency
$\nu_{\rm max}$ of the BLN shows a general trend to decrease with increasing
$S_{\rm a}$ in the upper banana, i.e.\ when kHz QPOs are not significantly
detected any more.
In the following we will discuss these and other results in more detail.

All the observations analyzed here belong to the same PCA gain epoch, 
PCA gain epoch 3, and therefore only small gain changes are expected. 
We used Crab observations taken close to the date of our observations to 
correct the colors and intensity of \1636 for these small instrumental 
effects.  The corrected color-color and hardness-intensity 
diagrams show significant long-term shifts of the atoll track.
Secular shifts in both the colors and the intensity in \1636 were previously 
reported by Prins \& van der Klis (1997) using EXOSAT 
observations. We find that
the shifts are more evident in the hardness-intensity diagram, where we 
observe two parallel branches in the upper banana, corresponding to two 
different intensity levels; it is also possible that the whole atoll track, 
and not just the upper banana branch, has shifted during the 1999 and part 
of the 1996 and 1998 observations, which correspond to the low-intensity 
levels (see Fig.~\ref{fig:fig1}).  Because the shift at the lower 
banana (left hand of the diagram) is along the track, we cannot clearly 
distinguish it from the movement of the source along the track.
However, the presence of a shift in this part of the diagram is confirmed
by the two-branch distribution of the kHz QPO frequencies versus intensity
diagram (see Fig.~\ref{fig:fig2}).
These shifts are much smaller but still visible in the color-color diagram,
where the upper banana branch has significantly moved towards lower soft 
colors (and probably higher hard colors) during the 1999 observations.  

Correlated to the long-term shifts of the atoll track, in the kHz QPO 
frequency vs.\ count rate diagram two parallel tracks are observed 
corresponding to two different flux levels (Fig.~\ref{fig:fig2}).
During the period of time spanned by our observations, about 2.5 years, 
\1636 is observed to move back and forth (once) between these two tracks; 
therefore for this source there seem to exist two preferred flux levels 
at which kHz QPOs are significantly detected. 
These two tracks tend to overlap each other if we plot the kHz QPO 
frequency vs.\ the colors or $S_{\rm a}$, although there are significant 
offsets still visible in the plots. These offsets are more noticeable 
in the plot vs.\ $S_{\rm a}$ or vs.\ the soft color, 
whereas they are less evident, although still present, in the plot vs.\ 
the hard color. 
Despite these shifts, the observed behavior is in general consistent with the 
hypothesis that kHz QPO frequencies depend only on the position of the source 
in the atoll track, irrespective of the motion of the track as a whole. 
However, the spread within the dataset P30053 (right hand panels of 
Fig.~\ref{fig:fig2}) indicates that there may be a more complex relation 
between kHz QPO frequencies and $S_{\rm a}$ or that the spline we chose does 
not go correctly through all these points. Another possibility is that there 
are shifts on a timescale of weeks/months of the atoll track in the color-color 
diagram along the soft color axis (not corresponding to evident intensity 
shifts, see Fig.~\ref{fig:fig2}) that are not clearly seen in the diagram 
because they occur along the banana branch.

The behavior observed in \1636 is similar to the secular
motion of the Z-track observed in the Z-sources Cyg X--2 (Kuulkers et al.\ 
1996), GX 5--1 (Kuulkers et al.\ 1994), GX 340+0 (Kuulkers \& van der Klis 
1996), and GX 17+2 (Wijnands et al.\ 1997b; Homan et al.\ 2002).
In these sources shifts in intensity in hardness-intensity diagrams are also 
reflected in (sometimes) much smaller, but still significant, shifts in the 
colors in color-color diagrams.
These variations have been interpreted in terms of a high inclination angle
of these sources, which would allow matter near the equatorial plane to 
modify the emission from the central region (e.g.\ Kuulkers et al.\ 1996).
However, as noted by van der Klis (2001), they could also be a consequence 
of a more general behavior of LMXB accretion, the same behavior that produces 
the parallel tracks in the kHz QPO frequencies vs.\ X-ray flux diagram.  
Indeed, our analysis confirms that secular motion of the 
atoll track produces the parallel tracks in \1636.

As a possible explanation for all this phenomenology,  
van der Klis (2001) proposed that kHz QPO frequencies, as well as other 
timing and spectral parameters, to first order depend on the disk inner 
radius, which is determined by the disk accretion rate normalized by its 
own long-term average, as could be expected in a radiative disk truncation 
scenario if luminosity responds, in part, `sluggishly' to disk accretion 
rate changes.  On the other hand, the X-ray luminosity is simply 
produced by the total accretion rate plus, possibly, nuclear burning.  This 
can be the case if part of the accreted matter escapes from the disk and flows 
in radially, at a rate that reflects a long term average of the disk accretion
rate, and some mechanism (e.g.\ radiation drag)
makes the inner disk radius dependent on the 
ratio of luminosity to disk accretion rate. In this scenario, the fact that 
the color-color diagram is also (but less) affected by the jump in the X-ray
flux (visible in the hardness-intensity diagram) indicates that the X-ray
spectral properties are mainly, but probably not completely, determined by 
the inner radius of the disk instead of by the total mass accretion rate.
In other words, changes in the flux level not affecting the frequencies of 
the kHz QPOs (i.e.\ the inner disk radius) may only slightly affect the 
position in the color-color diagram.

We studied the broad band (4 mHz - 2048 Hz frequency range) power spectra of 
\1636 as a function of the position of the source in the color-color 
diagram.  All power spectra are characteristic of the lower and upper 
banana states in atoll sources. Only three main components are observed
in these power spectra, the VLFN, the BLN, and one or two kHz QPOs.  
Since there is a shift in the upper part of the banana  
in the hardness-intensity diagram, due to the $\sim 20\%$ shift of the 
X-ray flux, for intervals 8 and 9 we extracted two different power spectra 
corresponding to the two different flux levels. We find that power 
spectra corresponding to the same position in the color-color diagram
but at different flux levels are quite similar to each 
other. The main difference is in the strength of the very low frequency
Lorentzian (Fig.~\ref{fig:fig4b}), which has an rms amplitude around $1\%$ 
at high flux level, and less than $\sim 0.7-0.8\%$ at low flux level.
A minor difference (if any) is in the fractional rms amplitude of the 
noise components (Figs.~\ref{fig:fig4} and \ref{fig:fig5}): the rms amplitudes 
are slightly ($\sim 2-2.7 \sigma$) higher for interval 9 low-flux than for
9 high-flux.  Note, however, that if we calculate the absolute rms amplitudes
(i.e.\ the fractional rms amplitude multiplied by the flux), these are
perfectly compatible between interval 9 low-flux and 9 high-flux.

We observe one or two kHz QPOs only in the lower part of the banana branch, 
up to $S_{\rm a} \sim 1.7$. 
For $\nu_{\rm low} \la 840$ Hz the upper peak is not significantly detected,
whereas we still detect the lower kHz QPO down to $\sim 780$ Hz.
The rms amplitude of the lower peak is approximately constant  
up to $\nu_{\rm low} \sim 850$ Hz (which is close to the frequency at 
which the upper peak begins to be significantly detected),
and then gradually decreases while $\nu_{\rm low}$ increases.
The rms amplitude of the upper peak is consistent with being constant
up to $\nu_{\rm low} \sim 900$ Hz, and then it significantly 
decreases when the lower peak frequency increases further 
(see Table~\ref{tab2}).  This is similar to what is observed in other atoll 
sources (see e.g.\ M\'endez et al.\ 2001; Di Salvo et al.\ 2001), 
where the rms amplitudes of the kHz QPOs remain constant or increase at 
low frequencies, and decrease at high frequencies.  
In 4U 1728--34 and 4U 1608--52 the amplitude of 
the upper kHz QPO decreases monotonically with frequency, whereas the rms 
amplitude of lower kHz QPO first increases, and then decreases with 
frequency.  In these two sources the rms amplitude of the upper kHz QPO is 
usually larger than that of the lower kHz QPO. On the contrary, in 
\1636 the rms amplitude of the lower peak is always larger than that 
of the upper peak (see Fig.~\ref{fig:fig3b}); this is probably because 
during these observations the kHz QPOs are at relatively high frequencies, 
where the lower peak becomes more prominent than the upper peak 
(cf.\ Fig.~3a in Di Salvo et al.\ 2001).  
Note also that for all these sources (including \1636) the rms amplitude 
of the kHz QPOs is nearly constant in the frequency range in which only one 
of them is detected.

We have also measured the frequency separation, $\Delta \nu$, between the  
kHz QPOs, when both of them were present simultaneously.  
In agreement with M\'endez et al.\ (1998a) we find that $\Delta \nu$ 
in \1636 is always significantly smaller than $290$ Hz, 
half the frequency of the quasi-coherent oscillations 
observed during type-I X-ray bursts in this source (see e.g.\ Strohmayer 
et al.\ 1998).
This is similar to what has been observed in another atoll source,  
4U 1728--34, where $\Delta \nu$ is always significantly smaller than the 
frequency of the nearly coherent oscillations observed in this source during 
type-I X-ray bursts, even at the lowest inferred mass accretion rate, when 
$\Delta \nu$ seems to reach its maximum value (M\'endez \& van der Klis 1999). 
In \1636, as well as in other similar sources, $\Delta \nu$ decreases
as the kHz QPO frequencies increase (van der Klis et al.\ 1997; M\'endez 
et al.\ 1998c; M\'endez \& van der Klis 1999; Ford et al.\ 1998; Homan 
et al.\ 2002; Jonker et al.\ 2002b; see also Fig.~\ref{fig:fig3b}). 
This trend cannot be explained by a simple beat frequency model, 
since in such model $\Delta \nu$ should be constant and equal to the 
neutron star spin frequency.  Among the different beat frequency models
that have been proposed (e.g.\ Strohmayer et al.\ 1996; Miller et al.\ 1998;
Campana 2000; Cui 2000), so far only the sonic-point beat frequency model 
can explain such an offset (Lamb \& Miller 2001; see, however, Jonker et al.\ 
2002a).  A decrease of the peak separation towards high, as well as towards 
low, frequencies is also predicted by the relativistic periastron precession 
model for the kHz QPOs (Stella \& Vietri 1999).
However, this model shows sometimes severe problems to fit the observed 
decrease, yielding reasonable fits only when allowing for highly eccentric 
orbits (Homan et al.\ 2002). 
The observational evidence that the kHz QPO frequencies are approximately
in a 2:3 frequency ratio led to the model based on parametric resonance
between the radial and vertical epicyclic frequencies in a relativistic 
accretion disk proposed by Klu\'zniak \& Abramowicz 2002 (see also
Abramowicz et al.\ 2002).
In this model the difference $\Delta \nu$ is not expected to be constant or 
equal to the neutron star spin frequency. However, if the kHz QPO frequencies  
stay in a constant ratio, $\nu_{\rm low} = 2/3\; \nu_{\rm high}$, then their 
difference $\Delta \nu$ should be correlated to the frequencies of the kHz 
QPOs, and not anticorrelated as is observed.  Note, in fact, that
there are evidences of a second peak at a value of $\sim 7/9$ in the 
distribution of the ratio $\nu_{\rm low} / \nu_{\rm high}$ in the case of 
Sco X--1 (Abramowicz et al.\ 2002); this high ratio might be due to the fact 
that the frequency difference decreases at high kHz QPO frequencies.

As in other atoll sources, the rms amplitude of the VLFN increases when 
the source moves towards the upper banana branch, although in \1636 it 
seems to saturate at high $S_{\rm a}$ values. 
We found that in some of the intervals the VLFN shows  
a shape that is more complex than a simple power law, because of the 
presence of a bump around a frequency between 20 and 30 mHz.
We fitted this bump using a Lorentzian, although it could also be due to 
a change in the slope of the power law describing the VLFN.  
This feature is not always significantly detected; it seems to be stronger
in intervals 4 to 7, with rms amplitudes between 1 and $1.4\%$, 
while it is much weaker in intervals 2 and 3, where its rms amplitude is 
$\sim 0.6-0.7\%$. In the intervals at the low-flux level the addition of this 
feature does not improve the fit, and the $90\%$ upper limits
on its rms amplitudes are $\sim 0.7-0.8\%$.
Deviations of the VLFN shape from a straight power law were
previously observed in some atoll sources; wiggles or bumps at a frequency 
around 0.1 Hz were particularly evident in the power spectra of the bright 
atoll sources GX 13+1, GX 9+9, and GX 9+1 (see Hasinger \& van der Klis 
1989).
Note that \1636 shows a mHz QPO at a quite stable frequency between 7 and 9
mHz, which is probably associated to some special mode of nuclear burning
at the neutron star surface (Revnivtsev et al.\ 2000; Yu \& van der Klis
2002), and there might be a relation between this mHz QPO and the bump
observed in the VLFN.  Indeed both features show similar rms amplitudes of 
$\sim 1\%$. 
However, while the mHz QPO 
is almost always present, with similar centroid frequencies (even in data 
separated by years), in the light curve of \1636, the VLFN bump is not equally 
stable in frequency and in rms amplitude. 
In particular, it does not seem to be present in the intervals at the 
low-flux level, while it is still detected with statistical significance 
in the corresponding intervals at high-flux level.

The last feature we want to discuss is the BLN, which is always significantly 
detected in the power spectra of \1636.  We fitted it with a Lorentzian, 
with a low quality factor (always less than 0.5); its rms amplitude 
generally decreases from $\sim 4\%$ to $\sim 2\%$ with increasing $S_{\rm a}$,
in line with the general behavior observed in atoll sources (e.g.\ 
Hasinger \& van der Klis 1989).  In agreement with previous results (Prins 
\& van der Klis 1997), we find that the frequency $\nu_{\rm max}$ of the BLN 
reaches quite high values, between 50 and 100 Hz, for $S_{\rm a} < 1.8$, among 
the highest ever observed for the BLN (typical values at high inferred 
accretion rates are 30--60 Hz, e.g.\ Belloni et al.\ 2002).  

\begin{figure}
\hbox{\hspace{0cm}\psfig{figure=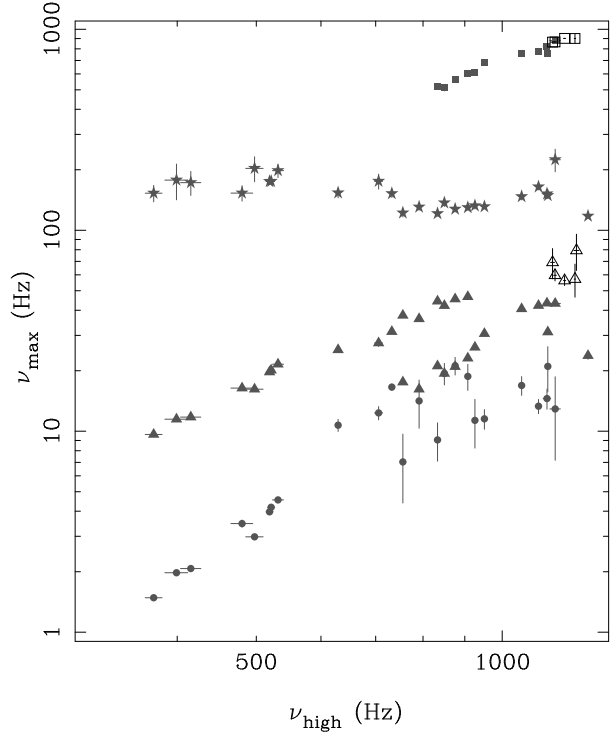,width=8.0cm}}
\caption[]{Characteristic frequencies ($\nu_{\rm max}$) 
of the several Lorentzians used to fit the power spectra of 4U 1728-34, 
4U 0614+09 (filled grey symbols, from van Straaten et al.\ 2001), and \1636 
(open black symbols, this paper) plotted versus the frequency of the upper 
kHz QPO. 
The filled circles mark the Lorentzian identified as the BLN, the filled
triangles the Lorentzian identified with the low-frequency QPO, the stars 
the hectohertz Lorentzian, and the squares the lower kHz QPO.  
In the case of \1636 the frequencies of the band limited noise are in the
region where the low-frequency QPO is expected on the basis of the observed
correlations, therefore we marked them with open triangles.}
\label{fig:fig6} 
\end{figure}

To compare our results on \1636 with those obtained for other sources
we plotted in Fig.~\ref{fig:fig6} the characteristic frequencies of the 
features observed in the \1636 power spectra (open black symbols), together 
with those observed in 4U 1728--34 and 4U 0614+09 (filled gray symbols, 
from van Straaten et al.\ 2002), versus the frequency of the upper kHz QPO, 
$\nu_{\rm high}$.  In this figure we can distinguish the following features, 
from the top to the bottom:
the lower kHz QPO (squares); the hectohertz QPO, a peaked noise component
with a quite constant frequency of $\sim 150$ Hz (stars); the LFQPO
(triangles), which weakens at high values of $\nu_{\rm high}$ and 
disappears for $\nu_{\rm high} \sim 900$ Hz; and the BLN (circles).
In 4U 1728--34 this last feature seems to become peaked for $\nu_{\rm high} 
\gta 900$ Hz: Di Salvo et al.\ (2001) interpreted these results supposing that
the noise component at low inferred accretion rates turns into a QPO at 
higher accretion rates, while what was called the LFQPO at low accretion 
rates gradually disappears at higher accretion rates.
Therefore in Fig.~\ref{fig:fig6} we marked the characteristic frequencies 
of this feature at high accretion rates with triangles and we will call 
it ``second LFQPO'' (note that the centroid frequency of this feature is 
approximately, but not exactly, half the frequency of the first LFQPO, so we 
cannot exclude it is a subharmonic of the first LFQPO). Simultaneously, another 
noise component appears, with a characteristic frequency of $\sim 7-10$ Hz; 
these points are marked with circles in Fig.~\ref{fig:fig6}.

As can be seen from this 
figure, the lower kHz QPO in \1636 follows the same correlation as that 
observed for the lower kHz QPO in 4U 1728--34 and 4U 0614+09.  The frequency 
of the BLN in \1636 follows a similar correlation to that observed for the 
LFQPO in 4U 1728--34 and 4U 0614+09.  Therefore, on the basis of these 
correlations, the BLN in \1636 can be tentatively identified with the 
feature that is called second LFQPO in 4U 1728--34 and 4U 0614+09. 
In the framework of the relativistic precession model, the LFQPO is interpreted 
as the Lense-Thirring precession frequency of the (slightly tilted) inner 
parts of the accretion disk, not coaligned with the equatorial plane 
of the neutron star, around the angular momentum axis of the neutron star 
(Stella \& Vietri 1998); according to the magnetospheric beat-frequency 
model, this would be the frequency difference between the Keplerian frequency 
at the magnetospheric radius
and the spin frequency of the neutron star (Psaltis et al.\ 1999b). In both 
these interpretations, the frequency of this feature is expected to increase 
with increasing inferred mass accretion rate, i.e.\ with decreasing inner 
radius of the disk.

However, from the correlation between the BLN frequency and $S_{\rm a}$ in 
\1636 (Fig.~\ref{fig:fig5}), we see that $\nu_{\rm max}$ tends to decrease at 
high values of $S_{\rm a}$ ($S_{\rm a} > 1.8$, note that the kHz QPOs in this 
source disappear for $S_{\rm a} \ge 1.7$).
Note, however, that we do not fit any hectohertz QPO in \1636 (the addition 
of this component is not statistically required), and this might 
affect our measures of the characteristic frequencies of the BLN if the
hectohertz QPO is present and not resolved. Indeed time variability seems much
weaker in \1636 than in 4U 1728--34; a comparison between the power spectra
of 4U 1728--34 and \1636 at similar kHz QPO frequencies (i.e. for 
$\nu_{\rm high} \simeq 1150$ Hz) shows that the total rms amplitude is higher 
in 4U 1728--34 than in \1636. However, it is not easy to see whether this is 
because some of the components are missing in the \1636 power spectrum,
or all the rms amplitudes are genuinely lower in \1636 than in 4U 1728--34.
Note that, for $\nu_{\rm high} \simeq 1150$ Hz, \1636 and 4U 1728--34 show
similar rms amplitudes of the upper and lower kHz QPO (respectively, 
$2.67\%$ and $6.5\%$ in \1636 and $3.68\%$ and $5.59\%$ in 
4U 1728--34) and of the VLFN ($2.01\%$ in \1636 and $3.39\%$ in 
4U 1728--34, cf.\ Di Salvo et al.\ 2001). Hence it is possible that some of
the time variability components are missing in \1636.

Indeed a decrease of the characteristic frequency of the BLN in the upper 
banana after the disappearance of the kHz QPOs may have been observed in 
some other similar sources.        
For instance, in 4U~1820--30 the cutoff frequency
of the cutoff power law used to describe the BLN was observed to decrease
from $\sim 13.7$ to $\sim 5.5$ Hz along the banana (Wijnands et al.\ 
1999). Further in the banana this noise component evolved 
first into a broad peaked noise and then into a QPO at a frequency of 
$\sim 7$ Hz (this behavior is, however, not observed in \1636, where this 
component is always quite broad).

Jumps to lower values in the frequency of the BLN have been observed in
4U 1728--34 and 4U 0614+09 (Ford \& van der Klis 1998; van Straaten et al.\ 
2000; Di Salvo et al.\ 2001; van Straaten et al.\ 2002b;  see also 
Fig.~\ref{fig:fig6}).  Also, the noise observed in the upper banana, when the 
kHz QPOs are no longer detected, is not easily identified in these sources.
This is because the power spectra of these sources are very complex 
(showing several noise components and QPOs) in the island state and lower
banana branch, while they become quite simple (showing only VLFN and
BLN) in the upper banana; it is not easy therefore to keep track of the 
evolution of all these components.  On the other hand, in the case of \1636 
the power spectra are much simpler than in other sources already in the 
lower banana branch, and this allows us to keep track of the evolution
of the BLN frequency.  If we tentatively identify in 4U 1728--34 and 
4U 0614+09 the BLN present in the upper banana with the second LFQPO (as we 
do in \1636), then in 4U 1728--34 at the transition to the upper banana, 
when the kHz QPOs are no longer detected and the VLFN becomes dominant,
the frequency of this feature significantly decreases from $\sim 43$ Hz to 
$\sim 33$ Hz and then to $\sim 21$ Hz, while in 4U 0614+09 it decreases 
from $\sim 31$ Hz to $\sim 24$ Hz (van Straaten et al.\ 2002b; see also 
van Straaten et al.\ 2000; Di Salvo et al.\ 2001).  
If this interpretation is correct, we can speculate that this feature 
evolves from a noise component to a QPO and then again to a noise component
in 4U~1728--34 and maybe in 4U 0614+09, while it is always quite 
broad in \1636.

A similar behavior may also have been observed in Z sources: a decrease of 
the HBO frequency with increasing $S_{\rm z}$ was in fact observed with high 
statistical significance in the Z source GX 17+2, when the kHz QPOs were 
still present (Wijnands et al.\ 1996, Homan et al.\ 2002). In that case, the 
fundamental frequency of the HBO increases from $\sim 20$ Hz to 
$\sim 60$ Hz for $S_{\rm z}$ increasing from --0.4 to 1.45, and then it 
significantly decreases from $\sim 60$ Hz to $\sim 48$ Hz for $S_{\rm z}$ 
increasing further up to $\sim 2$.  
This change in the behavior of the HBO frequency occurred when the source 
was in the Normal Branch, at high inferred mass accretion rates, while 
simultaneously the frequency of the kHz QPOs, which are thought to track the 
inner radius of the disk, continued to increase (Homan et al.\ 2002).

In the framework of the general relativistic precession model this behavior
may be explained in terms of the classical precession due to
oblateness of the neutron star (Morsink \& Stella 1999). However, the 
relation between the HBO and upper kHz QPO frequencies in GX 17+2 would 
require rather extreme neutron star parameters: a high neutron star spin 
frequency (500--700 Hz), a neutron star mass larger than $2 M_\odot$, and
a hard equation of state (see the discussion in Homan et al.\ 2002).
In the framework of the sonic point beat frequency model, this behavior
could be explained, for instance, assuming a two-flow model (Wijnands et al.\ 
1996), in which part of the accretion flow is radial outside the magnetosphere 
(to explain the HBO frequency decrease) and most of the radial flow 
falls back to the disk before the sonic point is reached (to explain the 
increase in the kHz QPO frequencies).  However, how this should work is
not clear, given that the magnetospheric radius and the sonic point radius 
should be close to each other.

\begin{acknowledgements}
We thank the referee, D. Barret, for useful suggestions.
TD likes to thank S. Migliari for helpful discussions.
This work was performed in the context of the research network "Accretion 
onto black holes, compact stars and protostars", funded by the European 
Commission under contract number ERB-FMRX-CT98-0195. 
This work was partially supported by the Netherlands Organization for 
Scientific Research (NWO).
\end{acknowledgements}


\begin{thebibliography}{}

\bibitem[]{}
Abramowicz, M. A., Bulik, T., Bursa, M., \& Klu\'zniak, W. 2002, A\&A, 
submitted (astro-ph/0206490)
\bibitem[]{}
Barret, D., \& Olive, J. F., 2002, ApJ, 576, 391
\bibitem[]{}
Belloni, T., Psaltis, D., \& van der Klis, M. 2002, ApJ, in press
(astro-ph/0202213)
\bibitem[]{}
Campana, S. 2000, ApJ, 534, L79
\bibitem[]{}
Cui, W. 2000, ApJ, 534, L, 31
\bibitem[]{}
Di Salvo, T., M\'endez, M., van der Klis, M., Ford, E., \& Robba, N. R.
2001, ApJ, 546, 1107
\bibitem[]{}
Ford, E. C., van der Klis, M., M\'endez, M., et al.\ 2000, ApJ, 537, 368
\bibitem[]{}
Ford, E., \& van der Klis, M. 1998, ApJ, 506, L39 
\bibitem[]{}
Gierlinski, M., \& Done, C. 2002, MNRAS, 331, L47
\bibitem[]{}
Giles, A. B., Hill, K. M., Strohmayer, T. E., \& Cummings, N. 2002, ApJ,
568, 222
\bibitem[]{}
Hasinger, G., \& van der Klis, M. 1989, A\&A, 225, 79
\bibitem[]{}
Hasinger, G., van der Klis, M., Ebisawa, K., Dotani, T., \& Mitsuda, K. 1990, 
A\&A, 235, 131
\bibitem[]{}
Hertz, P., Vaughan, B., Wood, K. S., et al.\ 1992, ApJ, 396, 201
\bibitem[]{}
Homan, J., van der Klis, M., Jonker, P.G., et al.\ 2002, ApJ, 568, 878
\bibitem[]{}
Jonker, P. G., Wijnands, R., van der Klis, M., et al.\ 1998, ApJ, 499, L191
\bibitem[]{}
Jonker, P. G., M\'endez, M., \& van der Klis, M. 2000, ApJ, 540, L29
\bibitem[]{}
Jonker, P. G., M\'endez, M., \& van der Klis, M. 2002a, MNRAS, 336, L1
\bibitem[]{}
Jonker, P. G., van der Klis, M., Homan, J., et al.\ 2002b, MNRAS, 333, 665
\bibitem[]{}
Klu\'zniak, W., \& Abramowicz, M. A. 2002, A\&A, submitted (astro-ph/0203314)
\bibitem[]{}
Kuulkers, E., van der Klis, M., Oosterbroek, T., et al.\ 1994, A\&A, 289, 795
\bibitem[]{}
Kuulkers, E., van der Klis, M., \& Vaughan, B.~A. 1996, A\&A, 311, 197
\bibitem[]{}
Kuulkers, E. \& van der Klis, M. 1996, A\&A, 314, 567
\bibitem[]{}
Lamb, F. K., \& Miller, M. C. 2001, ApJ, 554, 1210
\bibitem[]{}
M\'endez, M., van der Klis, M., van Paradijs, J., et al.\ 1997, ApJ, 485, L37
\bibitem[]{}
M\'endez, M., van der Klis, M., \& van Paradijs, J. 1998a, ApJ, 506, L117
\bibitem[]{}
M\'endez, M., van der Klis, M., van Paradijs, J., et al.\ 1998b, ApJ, 494, L65
\bibitem[]{}
M\'endez, M., van der Klis, M., Wijnands, R., et al.\ 1998c, ApJ, 505, L23
\bibitem[]{}
M\'endez, M., \& van der Klis, M. 1999, ApJ, 517, L51
\bibitem[]{}
M\'endez, M., van der Klis, M., Ford, E. C., Wijnands, R., \& van Paradijs, J.
1999, ApJ, 511, L49
\bibitem[]{}
M\'endez, M., van der Klis, M., \& Ford, E. C. 2001, ApJ, 561, 1016
\bibitem[]{}
Miller, M. C., Lamb, F. K., \& Psaltis, D. 1998, ApJ, 508, 791 
\bibitem[]{}
Miyamoto, S., Kimura, K., Kitamoto, S., Dotani, T., \& Ebisawa, K. 1991, ApJ, 
383, 784
\bibitem[]{}
Morsink, S. M., \& Stella, L. 1999, ApJ, 513, 827 
\bibitem[]{}
Muno, M. P., Remillard, R. A., \& Chakrabarty, D. 2002, ApJ, 568, L35
\bibitem[]{}
Nowak, M. A. 2000, MNRAS, 318, 361
\bibitem[]{}
Olive, J. F., Barret, D., Boirin, L., et al.\ 1998, A\&A 333, 942
\bibitem[]{}
Osherovich, V., \& Titarchuk, L. 1999, ApJ, 522, L113
\bibitem[]{}
Prins, S., \& van der Klis, M. 1997, A\&A, 319, 498
\bibitem[]{}
Psaltis, D., Belloni, T., \& van der Klis, M. 1999a, ApJ, 520, 262
\bibitem[]{}
Psaltis, D., Wijnands, R., Homan, J. et al.\ 1999b, ApJ, 520, 763
\bibitem[]{}
Revnivtsev, M., Churazov, E., Gilfanov, M., \& Sunyaev, R. 2001, A\&A, 372, 138
\bibitem[]{}
Stella, L., \& Vietri, M. 1998, ApJ, 492, L59
\bibitem[]{}
Stella, L., \& Vietri, M. 1999, Phys. Rev., 82, L17
\bibitem[]{}
Strohmayer, T. E., Zhang, W., Swank, J. H., White, N. E., \& Lapidus, I.
1998, ApJ, 498, L135
\bibitem[]{}
Strohmayer, T. E. 2001, in Astrophysical Sources of Gravitational Radiation 
for Ground-Based Detectors (astro-ph/0101160)
\bibitem[]{}
Strohmayer, T. E., Zhang, W., Swank, J. H., et al.\ 1996, ApJ, 469, L9 
\bibitem[]{}
Strohmayer, T. E., \& Markwardt, C. B., 2002, ApJ, 577, 337
\bibitem[]{}
Turner, M. J. L., \& Breedon, L. M. 1984, MNRAS, 208, 29
\bibitem[]{}
van der Klis, M. 2001, ApJ, 561, 943
\bibitem[]{}
van der Klis, M. 1995, in {\it X-Ray Binaries}, Lewin W. H. G., van Paradijs 
J., van den Heuvel E. P. J. eds., Cambridge Astrophysics Series, Cambridge
\bibitem[]{}
van der Klis, M. 2000, ARA\&A, 38, 717
\bibitem[]{}
van der Klis, M., Hasinger, G., Damen, E., et al.\ 1990, ApJ, 360, L19
\bibitem[]{}
van der Klis, M., Wijnands, R. A. D., Horne, K., \& Chen, W. 1997, 
ApJ, 481, L97
\bibitem[]{}
van der Klis, M., Jansen, F., van Paradijs, J., et al.\ 1985, Nature, 316, 225
\bibitem[]{}
van Paradijs, J., van der Klis, M., van Amerongen, S., et al.\ 
1990, A\&A, 234, 181
\bibitem[]{}
van Paradijs, J., Sztajno, M., Lewin, W. H. G., et al.\ 1986, MNRAS, 221, 617
\bibitem[]{}
van Straaten, S., van der Klis, M., \& M\'endez, M., 2002a, in preparation
\bibitem[]{}
van Straaten, S., van der Klis, M., Di Salvo, T., \& Belloni, T.
2002b, ApJ, 568, 912
\bibitem[]{}
van Straaten, S., Ford, E. C., van der Klis, M., M\'endez, M., \& Kaaret, P. 
2000, ApJ, 540, 1049
\bibitem[]{}
Wijnands, R., \& van der Klis, M. 1999, ApJ, 514, 939
\bibitem[]{}
Wijnands, R., van der Klis, M., \& Rijkhorst, E. J. 1999, ApJ, 512, L39
\bibitem[]{}
Wijnands, R. 2001, ApJ, 554, L59
\bibitem[]{}
Wijnands, R. A. D., van der Klis, M., van Paradijs, J., et al.\ 
1997a, ApJ, 479, L141
\bibitem[]{}
Wijnands, R., Homan, J., van der Klis, M., et al.\ 1997b, ApJ, 490, L157
\bibitem[]{}
Wijnands, R., van der Klis, M., Psaltis, D., et al.\ 1996, ApJ, 469, L5
\bibitem[]{}
Yu, W., \& van der Klis, M. 2002, ApJ, 567, L67
\bibitem[]{}
Zhang, W., Giles, A. B., Jahoda, K., et al.\ 1993, Proc. SPIE, 2006, 324
\bibitem[]{}
Zhang, W., Lapidus, I., White, N. E., \& Titarchuk, L. 1996, ApJ, 473, L135

\end{thebibliography}
\end{document}